\documentclass[a4paper,11pt]{article}
\usepackage{amsmath,amssymb,mathtools,cite,graphicx,fancyhdr}
\usepackage{bbold,color,mathpazo,authblk}
\usepackage{caption}
\graphicspath{{Graphs/}}

\begin{document}
\lhead{Solutions to PV equation through SUSY QM}
\rhead{Bermudez, Fern\'andez C., Negro}

\title{Solutions to the Painlev\'e V equation through \\
supersymmetric quantum mechanics}
\author[1]{David Bermudez\footnote{{\it email:} dbermudez@fis.cinvestav.mx}}
\author[1]{David J. Fern\'andez C.\footnote{{\it email:} david@fis.cinvestav.mx}}
\author[2]{Javier Negro\footnote{{\it email:} jnegro@fta.uva.es}}
\affil[1]{\textit{Departamento de F\'{\i}sica, Cinvestav, A.P. 14-740, 07000 Ciudad de M\'exico, Mexico}}
\affil[2]{\textit{Departamento de F\'{\i}sica Te\'orica, At\'omica y \'Optica. Universidad de Valladolid, 47011 Valladolid, Spain}}

\renewcommand\Authands{, and }

\date{}
\maketitle

\begin{abstract}
In this paper we shall use the algebraic method known as supersymmetric quantum mechanics (SUSY QM) to obtain solutions to the Painlev\'e V (PV) equation, a second-order non-linear ordinary differential equation. For this purpose, we will apply first the SUSY QM treatment to the radial oscillator. In addition, we will revisit the polynomial Heisenberg algebras (PHAs) and we will study the general systems ruled by them: for first-order PHAs we obtain the radial oscillator, while for third-order PHAs the potential will be determined by solutions to the PV equation. This connection allows us to introduce a simple technique for generating solutions of the PV equation expressed in terms of confluent hypergeometric functions. Finally, we will classify them into several solution hierarchies.\\

\end{abstract}

\section{Introduction}\label{intro}

Nowadays there is a growing interest in studying nonlinear phenomena and their corresponding description. This motivates us to look for the different relations which can be established between relevant physical subjects and nonlinear differential equations \cite{Sac91}. For example, the standard treatment for supersymmetric quantum mechanics (SUSY QM) leads to the Riccati equation \cite{AIS93,FF05}, which is the simplest non-linear first-order differential equation naturally associated with the search of eigenvalues for the Schr\"odinger Hamiltonian. Moreover, there are specific links for particular potentials, e.g., the SUSY partners of the free particle are connected with solutions of the KdV equation \cite{Lam80}. Is there something similar for systems different from the free particle?

The answer to this question turns out to be positive, concerning a connection which can be established between SUSY QM and Painlev\'e equations \cite{VS93,Adl94,Ber10,BF11a,Ber13}. Let us note that although these equations were discovered from strictly mathematical considerations, nowadays they are widely used to describe several physical phenomena \cite{AC92}. In particular, the Painlev\'e V (PV) equation appears in condensed matter \cite{Kan02}, electrodynamics \cite{Win92}, and solid state physics \cite{Lev92}. Moreover, this equation has attracted a lot of attention in the scientific community, thus leading to new studies about numerical solutions \cite{AY13}, geometric properties \cite{Sch02}, $q$-deformations \cite{Mas03}, discrete versions \cite{Mug07}, and B\"acklund transformations \cite{Sak05}, among others.

On the other hand, since its birth SUSY QM catalyzed the study of exactly-solvable Hamiltonians and gave a new insight into the algebraic structure characterizing them. Historically, the essence of SUSY QM was developed first as Darboux transformations in mathematical physics \cite{MS91} and later as the factorization method in quantum mechanics \cite{IH51,Mie84,junker}.

In this paper we are going to explore further the relation established between the SUSY partners of the radial oscillator and analytic solutions of the PV equation, using an approach similar to the one previously employed for studying the Painlev\'e IV (PIV) equation \cite{Ber13,BF11b,Ber12,BF12,BF13a}. This link has been found both in the context of dressing chains \cite{VS93,Adl94,DEK94} and in the framework of SUSY QM \cite{BF11a,ACIN00,FNN04,CFNN04,MN08}.

The key point of this connection is the following: the determination of general Schr\"odinger Hamiltonians having fourth-order differential ladder operators requires to find solutions to the PV equation. At the algebraic level this means that the corresponding systems are characterized by third-order polynomial deformations of the Heisenberg-Weyl algebra, called briefly polynomial Heisenberg algebras (PHA).

Reciprocally, if one wishes to obtain solutions to the PV equation first one looks for systems having fourth-order differential ladder operators; then, the corresponding solutions to the PV equation can be identified. It is worth to note that the first-order SUSY partners of the radial oscillator have associated natural fourth-order differential ladder operators, so that families of solutions to the PV equation can be easily obtained through this approach. Up to our knowledge, the first works in which it was realized the connection between PHA (called commutator representation in these papers) and Painlev\'e equations were \cite{ARS80} and \cite{Fla80}. Initially, both subjects were linked with first-order SUSY QM \cite{VS93,Adl94,DEK94}; later on, this relation was further explored for the higher-order case \cite{BF11a,Ber13,ACIN00,FNN04,CFNN04,MN08,SHC05,FC08,Mar09a,Mar09b}.

Let us remark that the need to avoid singularities in the new potential $V_k(x)$ and the requirement for the Hamiltonian $H_k$ to be Hermitian lead to some restrictions \cite{BF11a}: (i) first of all, the relevant transformation function has to be real, which implies that the associated factorization energy is also real; (ii) as a consequence, the spectrum of $H_k$ consists of two independent physical ladders, an infinite one departing from the ground state energy $E_0$ of $H_0$, plus a finite one with $k$ equidistant levels, all of which have to be placed below $E_0$. Regarding PV equation, these two restrictions imply that non-singular real solutions $w(z;a,b,c,d)$ can be obtained just for certain real parameters $a,b,c,d$.

From the spectral design point of view, however, it would be important to overcome restriction (ii) so that some (or all) steps of the finite ladder could be placed above $E_0$. In this way we would be able to manipulate not just the lowest part of the spectrum \cite{BF11a,FNN04,CFNN04}, but also the excited state levels, which would endow us with improved tools for spectral manipulation. 

In this work we are also going to show that this can be achieved if one relaxes restriction (i) by using complex transformation functions associated to real factorization energies \cite{AICD99}, which will permit us to generate complex solutions to PV equation associated with real parameters $a,b,c,d$, as it was done with the PIV equation \cite{BF11a,BF11b,Ber12,BF12,BF13a}. As a consequence, we will obtain complex SUSY partner potentials in this case. As far as we know, complex potentials with real energy spectra, obtained through complex transformation functions associated with real factorization energies, were first studied by Andrianov et al. \cite{AICD99}, and later by Fern\'andez et al. \cite{FMR03} for complex factorization {\it energies}. In this work we will also generate complex potentials with some \textit{complex energy levels}. Furthermore, the method we have developed to obtain solutions to the PV equation will be implemented as well.

In order to accomplish this, we will show that under certain conditions on the positions of the $k$ new levels and on the associated Schr\"odinger seed solutions, the combined results of SUSY QM (Section~\ref{capsusyqm}) and PHA (Sections~\ref{pha} and \ref{topha}) applied to the radial oscillator (Section \ref{SUSYRO}) lead to new solutions to the PV equation. Indeed, we will formulate and prove a {\it reduction theorem}, which imposes the restrictions on the transformation functions to reduce the natural ladder operators associated with $H_k$ from higher- to fourth-order (Section \ref{sectmaRO}). Then, we will study the properties of the different ladder operators, both the natural and the reduced ones, in order to analyze the consequences of the theorem. In particular, we will study the different types of PV solutions that can be obtained with this method (Sections \ref{solsPV}--\ref{cc}). 

Finally, after having obtained very general formulas for these PV solutions, a classification into several hierarchies will be introduced (Section \ref{hierarchies}) in order to compare them with other solutions which are spread in the literature \cite{FC08,WH03,CM08}. This classification is based upon the special functions used to explicitly write down the PV solutions. We will finish this paper with our conclusions (Section \ref{conclusions}).

\section{Supersymmetric quantum mechanics}
\label{capsusyqm}

The factorization method, intertwining technique, and SUSY QM are closely related and their names will be used indistinctly in this work to characterize a specific procedure through which it is possible to obtain new exactly-solvable quantum mechanical systems departing from known ones.

\subsection{First-order SUSY QM}
\label{secsusy1}
Let $H_0$ and $H_1$ be two Schr\"odinger Hamiltonians
\begin{equation}
	H_j = -\frac{1}{2}\frac{\text{d}^2}{\text{d}x^2} + V_j (x), \qquad j=0,1.
\end{equation}
For simplicity, we are taking {\it natural units} such that $\hbar=m=1$. Next, let us suppose the existence of a first-order differential operator $A_1^{+}$ that {\it intertwines} $H_0$ and $H_1$ in the way
\begin{equation}
	H_1 A_1^{+}=A_1^{+}H_0, \qquad A_1^{+}=\frac{1}{\sqrt{2}}\left[-\frac{\text{d}}{\text{d}x} + w_1 (x)\right],\label{entre}
\end{equation}
where the \emph{superpotential} $w_1 (x)$ is still to be determined. By plugging the explicit expressions for $H_0, \ H_1$ and $A_1^{+}$ into the intertwining relation \eqref{entre} and after some work we arrive to:
\begin{subequations}
	\begin{align}
		V_1(x) &= V_0(x) - w'_1(x,\epsilon),\label{Valfa}\\
		w'_1(x,\epsilon)+w^{2}_1(x,\epsilon) &= 2[V_0(x)-\epsilon]. \label{alfa}
	\end{align}\label{2alfas}
\end{subequations}
\vspace{-2mm}
\hspace{-1.5mm}If we define $u^{(0)}(x)$ such that $w_1(x,\epsilon)=u^{(0)'}/u^{(0)}$, then Equations \eqref{2alfas} are mapped into
\begin{subequations}
	\begin{align}
		V_1 &=  V_0- \left[\frac{u^{(0)'}}{u^{(0)}}\right]',\label{Vu}\\
		-\frac{1}{2}u^{(0)''} + V_0 u^{(0)} &= \epsilon u^{(0)},\label{ucero}
	\end{align}\end{subequations}
	i.e., $u^{(0)}$ is a solution of the initial stationary Schr\"odinger equation associated with $\epsilon$, although it might not fulfill any boundary condition at all.
	
	Starting from Equations \eqref{2alfas} we obtain that $H_0$ and $H_1$ can be factorized as
	\begin{equation}
		H_0 = A_1^{-} A^{+}_1+\epsilon,\qquad
		H_1 = A^{+}_1 A_1^{-}+\epsilon, \label{ffH}
	\end{equation}
	where
	\begin{equation}
		A_1^{-}\equiv(A^{+}_1)^{\dagger}=\frac{1}{\sqrt{2}}\left[\frac{\text{d}}{\text{d}x} + w_1(x,\epsilon)\right].\label{amam}
	\end{equation}
	
\begin{figure}
		\centering
		\includegraphics[scale=0.29]{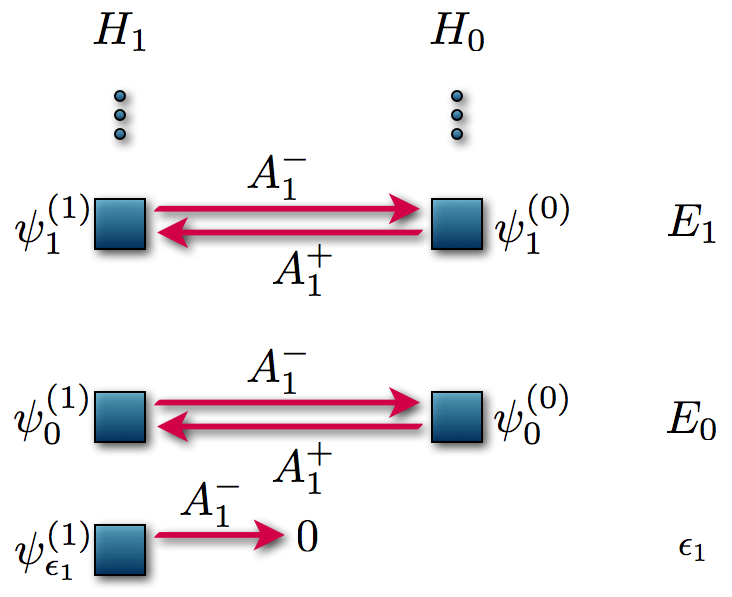}
		\caption{\small{Diagram of the first-order SUSY transformation. The final Hamiltonian $H_1$ has the same spectrum as the initial one $H_0$, but with a new level at the factorization energy $\epsilon_1$.}}
		\label{fig.susyqm1er}
	\end{figure}

	Let us assume that $V_0(x)$ is a solvable potential with normalized eigenfunctions $\psi^{(0)}_n (x)$ and eigenvalues such that Sp$(H_0)=\{E_n,n=0,1,2,\dots\}$. Besides, we know a non-singular solution $w_1(x,\epsilon)$ [or, equivalently, a solution $u_1^{(0)}(x)$ without zeroes] to the Riccati equation \eqref{alfa} [Schr\"odinger \eqref{ucero}] for $\epsilon=\epsilon_1 < E_0$, where $E_0$ is the ground state energy for $H_0$. Then, the potential $V_1(x)$ given in Equation \eqref{Valfa} [in \eqref{Vu}] becomes determined, and its normalized eigenfunctions are
	\begin{subequations}
		\begin{align}
			\psi^{(1)}_{\epsilon_1}(x) &\propto \exp\left(-\int^{x}_0 w_1(y,\epsilon_1)\text{d}y\right)=\frac{1}{u^{(0)}_1(x)},\label{psimathcal}\\
			\psi^{(1)}_{n}(x) &= \frac{A^{+}_{1} \psi_n^{(0)}(x)}{(E_n-\epsilon_1)^{1/2}},
		\end{align}\label{psis}
	\end{subequations}
	\hspace{-1.5mm}with eigenvalues such that Sp$(H_1)=\{\epsilon_1,E_n;n=0,1,\dots\}$. An scheme of the way the first-order SUSY transformation works, and the resulting spectrum, is shown in Figure \ref{fig.susyqm1er}. When this construction is implemented for a pair of well defined self-adjoint Hamiltonians $H_0, H_1$, we say that the supersymmetry is good or unbroken (see \cite{junker}).
	
	\subsection{Higher-order SUSY QM}
\begin{figure}\centering
		\includegraphics[scale=0.34]{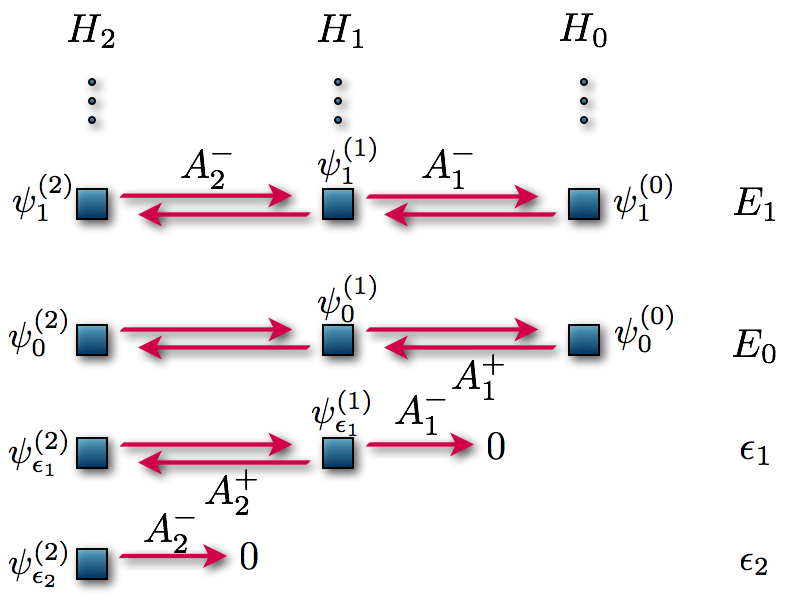}
		\caption{\small{Iteration of two first-order SUSY transformations applied to $H_0$ and $H_1$ using as transformation functions two mathematical eigenfunctions $u_1^{(0)}, \ u_2^{(0)}$ of $H_0$ with factorization energies $\epsilon_2 < \epsilon_1 < E_0$.}}\label{fig.susyqm2do}
	\end{figure}

	Let us iterate this technique, taking now $V_1(x)$ as the potential used to generate a new one $V_2(x)$ through another intertwining operator $A^{+}_2$ and a different factorization energy $\epsilon_2$, with the restriction $\epsilon_2<\epsilon_1 < E_0$ once again taken to avoid singularities in the new potentials and their eigenfunctions. The corresponding intertwining relation reads
	\begin{equation}
		H_2A^{+}_2=A^{+}_2H_1,
	\end{equation}
	\hspace{-1.5mm}which leads to equations similar to \eqref{2alfas} for $V_2$ and $w_2$:
	\begin{subequations}
		\begin{align}
			V_2(x) &= V_1(x) - w'_2(x,\epsilon_2),\label{Valfa2}\\
			w'_2(x,\epsilon_2)+w^{2}_2(x,\epsilon_2) &= 2[V_1(x)-\epsilon_2]. \label{alfa2}
		\end{align}\label{2alfas2}
	\end{subequations}
	\vspace{-2mm}
	\hspace{-1.5mm}In terms of $u^{(1)}_2(x)$ such that $w_2 (x,\epsilon_2)=u_2^{(1)}{}'(x)/u^{(1)}_2(x)$ we have
	\begin{equation}\label{Schro1}
		V_2  =  V_1 -\left[\frac{u_2^{(1)}{}'}{u^{(1)}_2}\right]',\qquad 
		-\frac{1}{2}u_2^{(1)}{}''+V_1 u^{(1)}_2 = \epsilon_2 u^{(1)}_2.
	\end{equation}
	Since the solution $u_2^{(1)}$ of the  Schr\"odinger equation \eqref{Schro1} is expressed in terms of two solutions $u_1^{(0)}, \ u_2^{(0)}$ of the initial Schr\"odinger equation associated to $\epsilon_1, \ \epsilon_2$ as
	\begin{equation}
		u_2^{(1)} \propto \frac{W(u_1^{(0)},u_2^{(0)})}{u_1^{(0)}} , 
	\end{equation}
	then the potential $V_2$ of Equation \eqref{Schro1} becomes:
	\begin{equation}
		V_2(x) = V_0(x) - \frac{\text{d}^2}{\text{d}x^2} \log W(u_1^{(0)},u_2^{(0)}).
	\end{equation}
	The eigenfunctions of $H_2$ are given by
	\begin{subequations}
		\begin{align}
			\psi^{(2)}_{\epsilon_2}(x) &\propto \exp\left(-\int^{x}_0 w_2(y,\epsilon_2)\text{d}y\right) \propto \frac{1}{u^{(1)}_2(x)} \propto \frac{u_1^{(0)}}{W(u_1^{(0)},u_2^{(0)})} , \label{psi22a}\\
			\psi^{(2)}_{\epsilon_1}(x) &= \frac{A^{+}_{2} \psi^{(1)}_{\epsilon_1}(x)}{(\epsilon_1-\epsilon_2)^{1/2}} \propto \frac{u_2^{(0)}}{W(u_1^{(0)},u_2^{(0)})},\\
			\psi^{(2)}_{n}(x) &= \frac{A^{+}_{2} \psi^{(1)}_{n}(x)}{(E_n-\epsilon_2)^{1/2}}= \frac{A^{+}_2A^{+}_1\psi^{(0)}_n(x)}{[(E_n-\epsilon_1)(E_n-\epsilon_2)]^{1/2}}\propto
			\frac{W(u_1^{(0)},u_2^{(0)},\psi_n^{(0)})}{W(u_1^{(0)},u_2^{(0)})},
		\end{align}
	\end{subequations}
	and the corresponding set of eigenvalues is Sp$(H_2)=\{\epsilon_2,\epsilon_1,E_n;n=0,1,\dots\}$. The scheme representing this transformation is shown now in Figure \ref{fig.susyqm2do}.
	
	This iterative process can be continued at will. Thus, let us assume that we know $k$ solutions $u_j^{(0)}(x)$ of the initial Schr\"odinger equation associated with $\epsilon_j$, $j=1,\dots,k$, where $\epsilon_{j+1}<\epsilon_{j}$. Therefore, we obtain a new solvable Hamiltonian $H_k$, whose potential reads
	\begin{equation}
		V_k(x)=V_0(x)-\frac{\text{d}^2}{\text{d}x^2}\log W(u_1^{(0)},u_2^{(0)},\dots ,u_k^{(0)}) .\label{Vk}
	\end{equation}
	
	The eigenfunctions of $H_k$ are given by
	\begin{subequations}
		\begin{align}
			\psi^{(k)}_{\epsilon_k}(x) &\propto \frac{W(u_1^{(0)},\dots ,u_{k-1}^{(0)})}{W(u_1^{(0)},\dots ,u_k^{(0)})},\\
			\psi^{(k)}_{\epsilon_{k-1}}(x)&\propto \frac{W(u_1^{(0)},\dots ,u_{k-2}^{(0)},u_k^{(0)})}{W(u_1^{(0)},\dots ,u_k^{(0)})},\label{psie}\\
			&\ \ \vdots \nonumber
		\end{align}
		\vspace{-12mm}
		\begin{align}
			\psi^{(k)}_{\epsilon_1}(x) &\propto \frac{W(u_2^{(0)},\dots ,u_{k}^{(0)})}{W(u_1^{(0)},\dots ,u_k^{(0)})},\\
			\psi^{(k)}_{n}(x) &\propto \frac{W(u_1^{(0)},\dots ,u_{k}^{(0)},\psi_{n¡}^{(0)})}{W(u_1^{(0)},\dots ,u_k^{(0)})}.\label{psin}
		\end{align}\label{psisk}
	\end{subequations}
	\hspace{-1.5mm}The corresponding eigenvalues are such that Sp$(H_k)=\{ \epsilon_j, E_n; j=k,\dots,1;$ $n=0,1,\dots \}$.
	
	By completeness, let us recall how the Hamiltonians $H_j$ are intertwined to each other
	\begin{equation}
		H_j A^{+}_j=A_j^{+}H_{j-1}, \qquad j=1,\dots ,k. \label{HAAH}
	\end{equation}
	Then, starting from $H_0$ we have generated a chain of factorized Hamiltonians in the way
	\begin{equation}
		H_j=A^{+}_jA_j^{-}+\epsilon_j,\qquad H_{j-1}=A^{-}_{j}A^{+}_{j}+\epsilon_{j}, \qquad j=1,\dots,k,
		\label{HAA}
	\end{equation}
	where the final potential $V_k(x)$ can be determined through Equation \eqref{Vk}.
	
	We must note now that we can define two $k$th-order differential operators as
	\begin{subequations}
		\begin{align}
			B_{k}^{+}&= A^{+}_k\dots A^{+}_1,\\
			B_{k}^{-}&= A^{-}_1\dots A^{-}_k,
		\end{align}\label{bs}
	\end{subequations}
	\hspace{-1.5mm}that intertwine the initial Hamiltonian $H_0$ with the final one $H_k$ as follows
	\begin{subequations}
		\begin{align}
			H_k B^{+}_k &= B^{+}_k H_0,\\
			B^{-}_k H_k &= H_0B^{-}_k.
		\end{align}\label{interB}
	\end{subequations}
	\hspace{-1.5mm}Therefore, the new Hamiltonian $H_k$ is determined by $k$ seed solutions $u_j^{(0)}$ of $H_0$ which are annihilated by $B_k^+$. Equations \eqref{psisk} and \eqref{interB} lead to
	\begin{subequations}
		\begin{align}
			B^{+}_k\psi^{(0)}_n &= [(E_n-\epsilon_1)\dots(E_n-\epsilon_k)]^{1/2}\psi^{(k)}_n,\label{Bdag2}\\
			B_k^{-}\psi^{(k)}_n &= [(E_n-\epsilon_1)\dots(E_n-\epsilon_k)]^{1/2}\psi^{(0)}_n.\label{Bdag3}
		\end{align}
	\end{subequations}
	
	These equations immediately lead to the higher-order SUSY QM \cite{AIS93,AICD95,BS97,FGN98,FHM98,Ros98a, Ros98b,FH99,BGBM99,MNR00}. In this treatment, the standard SUSY algebra with two generators $Q_1,\,Q_2$ \cite{Wit81},
	\begin{equation}
		[Q_i,H_{ss}]=0, \ \ \ \{Q_i,Q_j\}=\delta_{ij}H_{ss}, \ \ \ i,j=1,2,
		\label{Qis}
	\end{equation}
	can be realized from $B_k^{-}$ and $B^{+}_k$ through the definitions
	\begin{equation}
		Q^{-}=\left(\begin{array}{cc}
			0 &  0 \\
			B_k^{-} &  0 
		\end{array}\right),\
		Q^{+}=\left(\begin{array}{cc}
			0 &  B^{+}_k \\
			0 &  0 
		\end{array}\right),\
		H_{ss}\equiv\{Q^{-},Q^{+}\}=\left(\begin{array}{cc}
			B^{+}_kB_k^{-} &  0 \\
			0 &  B_k^{-}B^{+}_k 
		\end{array}\right),
		\label{Qis2}
	\end{equation}
	where $Q_1\equiv (Q^{+}+Q^{-})/\sqrt{2}$ and $Q_2\equiv i(Q^{-}-Q^{+})/\sqrt{2}$. Given that
	\begin{subequations}
		\begin{align}
			B^{+}_kB_k^{-}&=A^{+}_k\dots A^{+}_1 A_1^{-}\dots A_k^{-}=(H_k-\epsilon_1)\dots (H_k-\epsilon_k), \label{Bdag4}\\
			B_k^{-}B^{+}_k			&=A_1^{-}\dots A_k^{-} A^{+}_k\dots A^{+}_1=(H_0-\epsilon_1)\dots (H_0-\epsilon_k),\label{Bdag5}
		\end{align}\label{Bdag45}
	\end{subequations}
	\hspace{-1mm}it turns out that the SUSY generator $H_{ss}$ is a polynomial of degree $k$-th in the Hamiltonian $H^p_s$ that involves the two intertwined Hamiltonians $H_0$ and $H_k$,
	\begin{equation}
		H_{ss}=(H^p_s -\epsilon_1)\dots(H^p_s -\epsilon_k),
		\label{Hss}
	\end{equation}
	where
	\begin{equation}
		H^p_s=\left(\begin{array}{cc}
			H_k &  0 \\
			0 &  H_0 
		\end{array}\right).\\
		\label{Hps}
	\end{equation}
	
	Before applying the SUSY transformations to the particular potentials we are interested in, let us study first the polynomial deformations of the Heisenberg-Weyl algebra, which leads to Hamiltonians with very peculiar spectra. Moreover, for polynomial deformations of third-order we will arrive naturally to the PV equation. 
	
	\section{Polynomial Heisenberg algebras}\label{pha}
	
	Systems described by $(m-1)$-th order polynomial Heisenberg algebras (PHA) possess three generators, one of them (the Hamiltonian $H$) commutes with the other two (the ladder operators $\mathcal{L}_m^{\pm}$) in the same way as the harmonic oscillator Hamiltonian commutes with the creation and annihilation operators, but the commutator $[\mathcal{L}_m^{-},\mathcal{L}_m^{+}]$ is a polynomial in $H$, i.e.,
	\begin{subequations}
		\begin{align}
			[H,\mathcal{L}_m^{\pm}]&=\pm \mathcal{L}_m^{\pm}, \label{pha1} \\
			[\mathcal{L}_m^{-},\mathcal{L}_m^{+}]& \equiv N_m(H+1) - N_m(H)= P_{m-1} (H),\label{pha2}
		\end{align}\label{phaeqs}
	\end{subequations}
	where $P_{m-1}(H)$ and $N_m(H)\equiv \mathcal{L}_m^{+}\mathcal{L}_m^{-}$ are polynomials in $H$ of degrees $m-1$ and $m$ respectively. We can rewrite Equation \eqref{pha1} as
	\begin{equation}
		(H\mp 1)\mathcal{L}_m^\pm = \mathcal{L}_m^\pm H,
	\end{equation}
	which is equivalent to Equations \eqref{interB} with $H_0=H$, $H_k=H-1$ and $B_k^\pm=\mathcal{L}_k^\pm$ for $k=m$. These equations mark the connection between SUSY QM and PHA, i.e., we can obtain systems ruled by these algebras using SUSY transformations. Also, note that $N_m(H)$ is the analogous to the number operator of the harmonic oscillator, which can be factorized as
	\begin{equation}
		N_m(H)=\prod_{i=1}^{m}(H-\varepsilon_{i}),\label{facN}
	\end{equation}
	with $\varepsilon_i$ being the energies associated with the {\it extremal states}. 
	
	The PHA can be realized through $m$th-order differential ladder operators, but the degree $m-1$ of the polynomial characterizing the deformation (see commutator in Equation \eqref{pha2}) is what defines the order of the PHA. From now on, we will suppose that $H$ has Schr\"odinger form and $\mathcal{L}_m^{\pm}$ are $m$th-order differential ladder operators.
	
\begin{figure}
		\centering
		\includegraphics[scale=0.24]{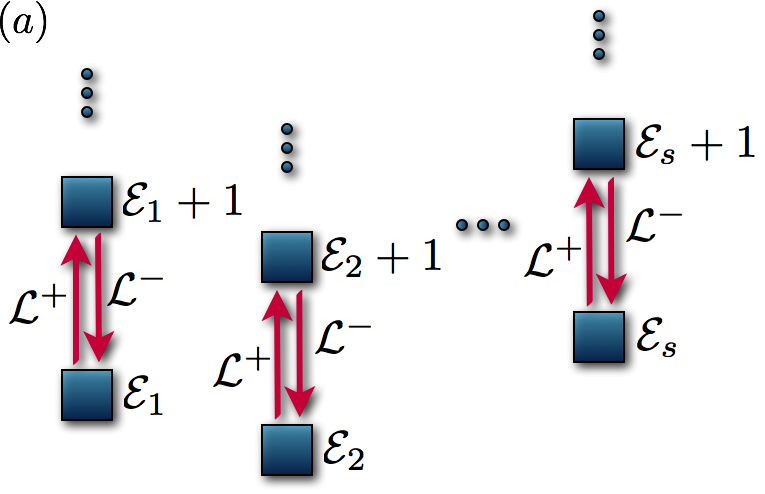}\hspace{10mm}
		\includegraphics[scale=0.24]{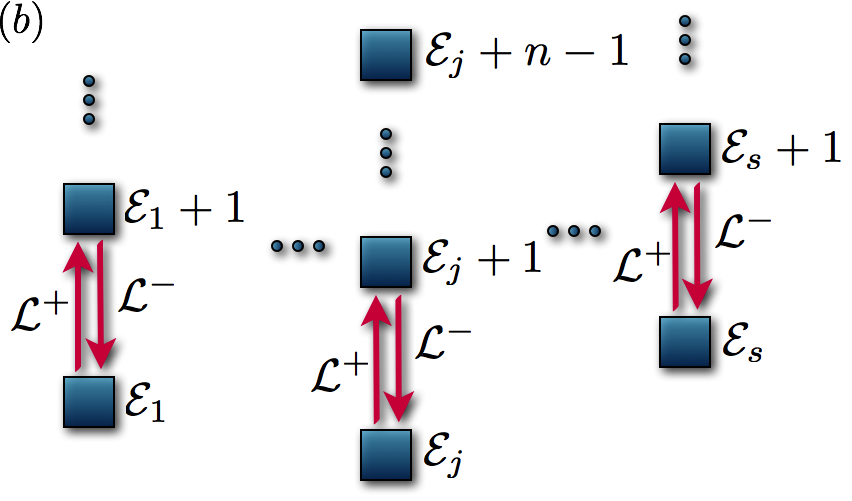}
		\caption{\small{In $(a)$ we show the spectrum for a Hamiltonian with $s$ physical extremal states, each one of them having associated, in general, one infinite ladder. In $(b)$ we show an spectrum where $\psi_{\varepsilon_j}$ fulfills condition \eqref{finita} and thus the system has $s-1$ infinite ladders and one finite (the $j$-th).}}\label{fig.pha1b}
	\end{figure}

	The algebra generated by $\{H,\mathcal{L}_m^{-},\mathcal{L}_m^{+}\}$ supplies information about the spectrum of $H$, $\text{Sp}(H)$ \cite{ACIN00,FH99,DEK92}. In fact, let us consider the $m$th-dimensional solution space of the differential equation $\mathcal{L}_m^{-}\psi=0$, called the {\it kernel} of $\mathcal{L}_m^{-}$ and denoted as $\mathcal{K}_{\mathcal{L}_m^{-}}$. Then
	\begin{equation}
		\mathcal{L}_m^{+}\mathcal{L}_m^{-}\psi = \prod_{i=1}^{m}(H-\varepsilon_i )\psi=0.
	\end{equation}
	Since $\mathcal{K}_{\mathcal{L}_m^{-}}$ is invariant with respect to $H$, then it is natural to select the formal eigenfunctions of $H$ as basis for the solution space, i.e., $H\psi_{\varepsilon_i}=\varepsilon_i \psi_{\varepsilon_i}$. Therefore, $\psi_{\varepsilon_i}$ become the extremal states of $m$ ladders of \textit{formal} or \textit{mathematical} eigenfunctions with spacing $\Delta E=1$ that start from $\varepsilon_i$. Let $s$ be the number of those states with physical significance, $\{ \psi_{\varepsilon_i} ;i=1,\dots ,s \}$; then, using $\mathcal{L}_m^{+}$ we can construct from them $s$ physical energy ladders of infinite length, as shown in Figure \ref{fig.pha1b}$(a)$. There also exist cases where $H$ has a degenerate structure, then the basis of $\mathcal{K}_{\mathcal{L}_m^{-}}$ will induce a Jordan cell in its matrix representation and it will be non-diagonalizable. However, in this work we will not deal with such degenerate cases.
	
	It is possible that, for a ladder starting from $\varepsilon_j$, there exists an integer $n\in N$ such that
	\begin{equation}
		(\mathcal{L}_m^{+})^{n-1}\psi_{\varepsilon_j} \neq 0,\qquad
		(\mathcal{L}_m^{+})^{n}\psi_{\varepsilon_j} = 0.
		\label{finita}
	\end{equation}
	Then, if we analyze the expression $\mathcal{L}_m^{-}(\mathcal{L}_m^{+})^{n}\psi_{\varepsilon_j} = 0$ we can see that another root of Equation~\eqref{facN} must fulfill $\varepsilon_k = \varepsilon_j +n$, where $k\in \{s+1,\dots ,m\}$ and $\ j\in \{1,\dots,s \}$. Therefore, in this case Sp$(H)$ will contain $s-1$ infinite ladders and a finite one with length $n$, that starts in $\varepsilon_j$ and finishes at $\varepsilon_j +n-1$, as it is shown in Figure \ref{fig.pha1b}$(b)$.
	
	In conclusion, the spectra of systems described by an $(m-1)$th-order PHA can have at most $m$ infinite ladders. Note that the standard annihilation and creation operators $a^\pm$ for the harmonic oscillator together with its Hamiltonian satisfy Equations (\ref{phaeqs}--\ref{facN}) for $m=1$. Moreover, a higher-order algebra with odd $m$ can be constructed simply by taking $\mathcal{L}_m^{-}=a^{-}Q(H)$, $\mathcal{L}_m^{+}=Q(H)a^{+}$, where $Q(H)$ is a real polynomial of $H$ \cite{DGRS99}. These deformations are called \emph{reducible} and in this context they become trivial, since our system already has operators $a^{\pm}$ fulfilling a lower-order algebra.
	
	Now, it is important to identify the general systems ruled by these PHA. It has been seen that the difficulties in this study grow with the order of the algebra: for zeroth- and first-order PHA, the systems turn out to be the harmonic and radial oscillators, respectively \cite{DEK92,Fer84D,Adl93,SRK97}. On the other hand, for second- and third-order PHA, the determination of the potentials reduces to find solutions to the Painlev\'e IV and V equations \cite{WH03,Adl93}. This connection can be used in reverse order, i.e., first, look for systems which are certainly described by PHA and then find solutions to the Painlev\'e equations. This solution method was successfully used for the Painlev\'e IV equation \cite{Ber10,BF11a,BF11b,BF12,Ber12,BF13a,Ber13}. In this work we will expand it to the PV equation. In order to do that, let us see next the way in which this connection appears for third-order PHA and PV equation.
	
	\section{Third-order PHA: fourth-order ladder operators.}\label{topha}
\begin{figure}
		\begin{center}
			\includegraphics[scale=0.3]{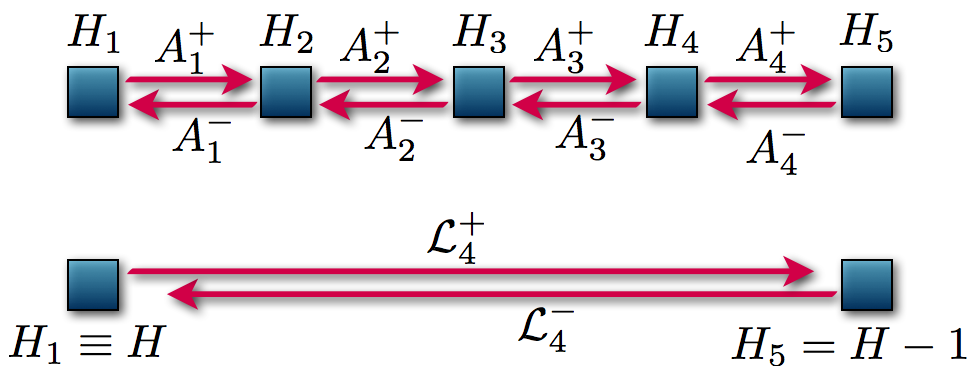}
		\end{center}
		\vspace{-5mm}
		\caption{\small{Diagram representing the two equivalent SUSY transformations. Above: the four-step first-order transformation induced by $A_1^{\pm}$, $A_2^{\pm}$, $A_3^{\pm}$ and $A_4^\pm$. Below: the direct transformation achieved through the fourth-order operators $\mathcal{L}_4^{\pm}$.}}
		\label{diasusyRO}
	\end{figure}

	Let us suppose that $\mathcal{L}_4^{\pm}$ are fourth-order ladder operators, which are factorized as follows:
	\begin{subequations}
		\begin{align}
			\mathcal{L}_4^{+}&=A_4^+A_{3}^{+}A_{2}^{+}A_{1}^{+}=
			\frac{1}{4}\left(\frac{\text{d}}{\text{d}x}-f_4\right)\left(\frac{\text{d}}{\text{d}x}-f_3\right)\left(\frac{\text{d}}{\text{d}x}-f_2\right)\left(\frac{\text{d}}{\text{d}x}-f_1\right),\\
			\mathcal{L}_4^{-}&=A_{1}^{-}A_{2}^{-}A_{3}^{-}A_4^-=
			\frac{1}{4}\left(-\frac{\text{d}}{\text{d}x}-f_1\right)\left(-\frac{\text{d}}{\text{d}x}-f_2\right)\left(-\frac{\text{d}}{\text{d}x}-f_3\right)\left(-\frac{\text{d}}{\text{d}x}-f_4\right)\label{lmenosRO}.
		\end{align}\label{elesRO}
	\end{subequations}
	\hspace{-1mm}As we saw in Section 3, these are particular cases of the factorization in Equation \eqref{bs}. Now, let us build a closed-chain so that each pair of operators $A_j^-$, $A_j^+$ intertwines two Schr\"odinger Hamiltonians $H_j$ and $H_{j+1}$ in the way \cite{Adl94}
	\begin{equation}
		H_{j+1}A^{+}_j =A^{+}_{j}H_{j},\qquad H_{j}A^{-}_j =A^{-}_{j}H_{j+1},
	\end{equation}
	where $j=1,2,3,4$. This leads to the following factorizations of the Hamiltonians
	\begin{subequations}\label{fac-folo}
		\begin{align}
			H_{1}&=A_{1}^{-}A_{1}^{+}+\epsilon_{1},\\
			H_{2}&=A_{1}^{+}A_1^{-}+\epsilon_1=A_{2}^{-}A_{2}^{+}+\epsilon_{2},\\
			H_{3}&=A_{2}^{+}A_2^{-}+\epsilon_2=A_3^{-}A_{3}^{+}+\epsilon_{3},\\
			H_{4}&=A_{3}^{+}A_3^{-}+\epsilon_3=A_4^{-}A_{4}^{+}+\epsilon_{4},\\
			H_{5}&=A_{4}^{+}A_4^{-}+\epsilon_4.
		\end{align}
	\end{subequations}
	To accomplish the closed-chain we need the closure condition given by
	\begin{equation}
		H_5=H_1-1\equiv H-1.\label{cloconRO}
	\end{equation}
	In Figure~\ref{diasusyRO} we show a diagram representing the transformations and the closure relation.
	
	By making the corresponding operator products of Equations \eqref{fac-folo} we obtain the following systems of equations
	\begin{subequations}
		\begin{align}
			f_1'+f_2'&=f_1^2-f_2^2+2(\epsilon_1-\epsilon_2),\label{f1RO}\\
			f_2'+f_3'&=f_2^2-f_3^2+2(\epsilon_2-\epsilon_3),\label{f2RO}\\
			f_3'+f_4'&=f_3^2-f_4^2+2(\epsilon_3-\epsilon_4),\label{f3RO}\\
			f_4'+f_1'&=f_4^2-f_1^2+2(\epsilon_4-\epsilon_1+1).\label{f4RO}
		\end{align}\label{efesRO}
	\end{subequations}
	In order to simplify notation let us make $\alpha_1=\epsilon_1-\epsilon_2$, $\alpha_2=\epsilon_2-\epsilon_3$, $\alpha_3=\epsilon_3-\epsilon_4$, and $\alpha_4=\epsilon_4-\epsilon_1+1$. If all Equations~\eqref{efesRO} are added it is obtained:
	\begin{equation}
		f_1+f_2+f_3+f_4=x.\label{sumas}
	\end{equation}
	Since the system is over-determined, a constrain $A$ is going to be employed
	\begin{equation}
		f_1^2-f_2^2+f_3^2-f_4^2=\alpha_4-\alpha_3+\alpha_2-\alpha_1\equiv A.\label{restr1}
	\end{equation}
	Taking into account \eqref{sumas} and \eqref{restr1} the system of Equations \eqref{efesRO} can be reduced to a non-linear second-order one. Thus, let us denote
	\begin{equation}
		g\equiv -f_1-f_2,\qquad p\equiv f_1-f_2,\qquad q\equiv f_2+f_3,
		\label{gpq}
	\end{equation}
	so that Equations~\eqref{f1RO} and \eqref{f2RO} can be written as
	\begin{equation}
		g'=gp-2\alpha_1,\qquad q' =-q(q+g+p)+2\alpha_2,\label{gqp}
	\end{equation}
	and the restriction in Equation \eqref{restr1} becomes
	\begin{equation}
		x p + (g+x)(2q-x)=A, \label{restr2}
	\end{equation}
	i.e., we have now the system of three Equations \eqref{gqp} and \eqref{restr2}. If we define $t\equiv 2q-x$ and then solve $p$ from Equation~\eqref{restr2} we arrive at
	\begin{equation}
		p=\frac{1}{x}[A-(x+g)t].\label{pgt}
	\end{equation}
	By substituting this expression into both Equations~\eqref{gqp}, a system of two equations is obtained
	\begin{equation}
		g' =\frac{g}{x}[A-(x+g)t]-2\alpha_1,\qquad t'=(t+x)\left(\frac{gt-A}{x}+\frac{t-x}{2}-g\right)+4\alpha_2-1.\label{gh}
	\end{equation}
	By defining now two new functions $w$ and $v$ as
	\begin{equation}
		xt(x)  = v(x^2),\qquad g(x)  = \frac{x}{w(x^2)-1}, \label{vw}
	\end{equation}
	and making the change $x^2\rightarrow z$, the system of equations \eqref{gh} becomes
	\begin{subequations}
		\begin{align}
			v' & = \left(\frac{v^2}{4z}-\frac{z}{4}\right)\left(\frac{w+1}{w-1}\right)+(1-A)\frac{v}{2z}+2\alpha_2-\frac{A}2 - \frac12,\label{vw1}\\
			w' & = \frac{\alpha_1}{z}(w-1)^2+\frac{(1-A)}{2z}(w-1)+\frac{vw}{2z},\label{vw2}
		\end{align}
	\end{subequations}
	where now the derivatives are with respect to $z$. We clear then $v$ from Equation~\eqref{vw2}, derive this result with respect to $z$ and substitute $v$ and $v'$ in Equation~\eqref{vw1}. After a long calculation we obtain finally a single equation for $w$ given by
	\begin{equation}
		w''=\left(\frac{1}{2w}+\frac{1}{w-1}\right)(w')^2-\frac{w'}{z}+\frac{(w-1)^2}{z^2}\left(aw+\frac{b}{w}\right)+c\frac{w}{z}+d\frac{w(w+1)}{w-1},\label{PV}
	\end{equation}
	which is the Painlev\'e V equation with parameters
	\begin{equation}
		a=\frac{\alpha_1^2}{2},\qquad b=-\frac{\alpha_3^2}{2},\qquad c=\frac{\alpha_2-\alpha_4}{2},\qquad d=-\frac{1}{8}.\label{paraPV}
	\end{equation}
	
	The spectrum of $H$ could contain four independent equidistant energy ladders starting from the following extremal states \cite{CFNN04}:
	\begin{subequations}
		\begin{align}
			\psi_{\varepsilon_1} \propto &\left[\frac{h}{2}\left(\frac{g'}{2g}-\frac{h'}{2h}-\frac{x}{2}-\frac{\alpha_1}{g}\right)-\alpha_1-\alpha_2-\frac{\alpha_3}{2}\right] 
			\exp\left[\int\left(\frac{g'}{2g}+\frac{g}{2}-\frac{\alpha_1}{g}\right)\text{d}x\right],\\
			\psi_{\varepsilon_2} \propto & \left[\frac{h}{2}\left(\frac{g'}{2g}-\frac{h'}{2h}-\frac{x}{2}+\frac{\alpha_1}{g}\right)-\alpha_2-\frac{\alpha_3}{2}\right]
			\exp\left[\int\left(\frac{g'}{2g}+\frac{g}{2}+\frac{\alpha_1}{g}\right)\text{d}x\right],\\
			\psi_{\varepsilon_3} \propto & \exp\left[\int\left(\frac{h'}{2h}+\frac{h}{2}-\frac{\alpha_3}{h}\right)\text{d}x\right],\label{extRO3}\\
			\psi_{\varepsilon_4} \propto & \exp\left[\int\left(\frac{h'}{2h}+\frac{h}{2}+\frac{\alpha_3}{h}\right)\text{d}x\right],\label{extRO4}
		\end{align}\label{extremalRO}
	\end{subequations}
	\hspace{-1mm}where
	\begin{equation}
		h(x)=-x-g(x).\label{hdeg}
	\end{equation}
	Note that the number operator $N_4(H)$ for this system will be a fourth-degree polynomial:
	\begin{equation}
		N_4(H)=(H-\varepsilon_1)(H-\varepsilon_2)(H-\varepsilon_3)(H-\varepsilon_4).
	\end{equation}
	From Equations \eqref{Bdag45} and (\ref{elesRO}--\ref{cloconRO}), the energies of the extremal states can be obtained in terms of the factorization energies as
	\begin{equation}
		\varepsilon_1=\epsilon_1+1,\qquad \varepsilon_2=\epsilon_2+1,\qquad \varepsilon_3=\epsilon_3+1,\qquad \varepsilon_4=\epsilon_4+1.\label{eeps}
	\end{equation}
	Thus, if we have a solution $w$ of the PV equation~\eqref{PV}, consistent with the parameters of Equation~\eqref{paraPV}, we obtain a system characterized by a third-order PHA. In fact, once we get $w(z)$ we can obtain $v(z)$ from Equation~\eqref{vw2}, then from these two we can obtain $t,g$ using Equations~\eqref{vw}. After that we obtain also $p$ from Equation \eqref{pgt} and $q$ from Equation \eqref{restr2}. Finally, we can go back to $f_1$, $f_2$, $f_3$, $f_4$ employing Equation~\eqref{sumas} and definitions \eqref{gpq}.
	
	Conversely, if a system with fourth-order ladder operators is found, then it is possible to design a mechanism for generating solutions to the PV equation, similar to the one implemented for the PIV equation \cite{Ber13,Ber12,BF12}. The key point is to identify once again the extremal states of our system. Then, from Equations~\eqref{extRO3} and \eqref{extRO4} it is straightforward to show that
	\begin{align}
		h(x)&=\frac{2\alpha_3}{\left[\ln\left(\frac{\psi_{\varepsilon_4}}{\psi_{\varepsilon_3}}\right)\right]'}=\left\{\ln\left[W(\psi_{\varepsilon_3},\psi_{\varepsilon_4})\right]\right\}'.\label{ache}
	\end{align}
	Therefore, we obtain an expression for $g(x)$ given by
	\begin{align}
		g(x)&=-x-h(x)=-x-\left\{\ln\left[W(\psi_{\varepsilon_3},\psi_{\varepsilon_4})\right]\right\}'.\label{gdex}
	\end{align}
	Note that $g(x)$ is related with the solution $w(z)$ of the PV equation through
	\begin{equation}\label{wdez}
		w(z)=1+\frac{z^{1/2}}{g(z^{1/2})}.
	\end{equation}
	Thus, we have introduced a simple recipe to generate solutions to the PV equation, based on the identification of the extremal states for systems ruled by third-order PHA, which have differential ladder operators of fourth order.
	
	Before we finish this section, let us point out that in general the solutions $w(z)$ to the PV equation will depend on the four parameters involved, i.e., $w=w(a,b,c,d;z)$. In addition, since the general solution of the PV equation cannot be written in terms of standard special functions, these solutions are called {\it Painlev\'e V transcendents}. Nevertheless, for specific values of the parameters $a,b,c,d$ some of them can actually be expressed in terms of standard special functions. Solutions of this type which are found in the literature can be given in terms of rational functions, Laguerre or Hermite polynomials, Weber or Bessel functions, among others. In this paper we are going to obtain explicit solutions to the PV equation, initially expressed in terms of the confluent hypergeometric function. Then, we will derive some of the solution families which are related with other special functions, called {\it solution hierarchies}, in order to compare our results with the ones existent in the literature (see e.g. \cite{BF13a}).
	
	\section{SUSY partners of the radial oscillator}\label{SUSYRO}
	In this section we apply the SUSY QM of $k$-th order to the radial oscillator Hamiltonian $H_\ell$, which is given by
	\begin{equation}
		H_\ell=-\frac{1}{2}\frac{\text{d}^2}{\text{d}x^2}+\frac{x^2}{8}+\frac{\ell(\ell+1)}{2x^2}, \qquad \ell\geq -\frac{1}{2}, \qquad x\geq 0, \label{Hell}
	\end{equation}
	and whose eigenfunctions $\psi_{n\ell}$ and associated eigenvalues $E_{n\ell}$ satisfy:
	\begin{equation}
		H_\ell \psi_{n\ell} = E_{n\ell}\psi_{n\ell}, \qquad \psi_{n\ell}(0)=\psi_{n\ell}(\infty)=0.
		\label{eigenfun}
	\end{equation}
	The subscript $\ell$ was added to denote the Hamiltonian dependence on the angular momentum index. Before doing the SUSY treatment, however, let us determine the spectrum of $H_\ell$ through the usual factorization method \cite{CFNN04,FND96,CR08}. 
	
	The radial oscillator Hamiltonian can be factorized in four different ways, the first two are:
	\begin{equation}
		H_\ell=a_\ell^- a_\ell^+ +\frac{\ell}{2}-\frac{1}{4} = a_{\ell+1}^+a_{\ell+1}^-+\frac{\ell}{2}+\frac{3}{4},
	\end{equation}
	with
	\begin{equation}
		a_\ell^\pm \equiv \frac{1}{\sqrt{2}}\left(\mp\frac{\text{d}}{\text{d}x}-\frac{\ell}{x}+\frac{x}{2}\right).
	\end{equation}
	The commutator of the last two operators is
	\begin{equation}
		[a_\ell^-,a_\ell^+]=\frac{\ell}{x^2}+\frac{1}{2},
		\label{conmdef}
	\end{equation}
	from which we conclude that $a_\ell^\pm$ are neither ladder operators nor shift-operators, but rather a mixture of both, since they change the energy $E$ as well as the angular momentum index $\ell$. An alternative definition of the commutator would be
	\begin{equation}
		[a^-,a^+]_\ell \equiv a_\ell^-a_\ell^+ - a_{\ell+1}^+ a_{\ell+1}^- = 1,
	\end{equation}
	which depends on the index $\ell$, i.e., it would be different for each Hamiltonian. However, in this work we will stick to the standard definition of Equation \eqref{conmdef}.
	
	Operators $a_\ell^\pm$ intertwine $H_\ell$ with $H_{\ell-1}$, creating a hierarchy of Hamiltonians with different $\ell$:
	\begin{equation}
		H_\ell a_\ell^- = a_\ell^-\left(H_{\ell-1}-\frac{1}{2}\right),\qquad H_{\ell-1} a_\ell^+ = a_\ell^+\left(H_{\ell}+\frac{1}{2}\right).\label{interRO}
	\end{equation}
	Now, let us take an eigenfunction $\psi_{n\ell}(x)$ of $H_\ell$ with eigenvalue $E_{n\ell}$, as defined in Equation \eqref{eigenfun}. Then, from Equations~\eqref{interRO} it is obtained that
	\begin{equation}
		H_{\ell+1} (a_{\ell+1}^- \psi_{n\ell}) = \left(E_{n\ell}-\frac{1}{2}\right)(a_{\ell+1}^-\psi_{n\ell}),\qquad
		H_{\ell-1} (a_{\ell}^+ \psi_{n\ell}) = \left(E_{n\ell}+\frac{1}{2}\right)(a_{\ell}^+\psi_{n\ell}).
	\end{equation}
	
	\begin{figure}
		\centering
		\includegraphics[scale=0.34]{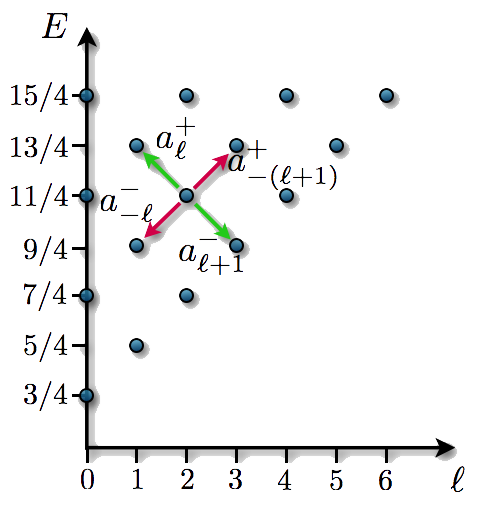}
		\includegraphics[scale=0.34]{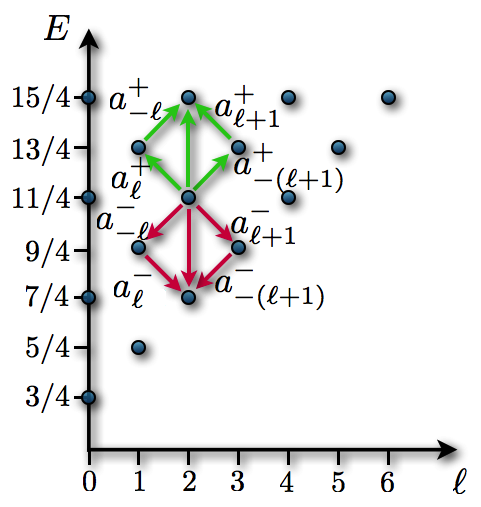}
		\caption{\small{Diagram of the action of the first-order shift operators $a_\ell^\pm$ and $a_{-\ell}^\pm$ (left). The horizontal axis represents the angular momentum index $\ell$ and the vertical one the energy. The joint action of two appropriate shift operators leads to the second-order ladder operators $b_\ell^\pm$ (right).}}\label{diaRO}
	\end{figure}
	On the other hand, the substitution $\ell\rightarrow -(\ell+1)$ produces the other two factorizations, as well as corresponding changes in the equations. For the factorizations we have
	\begin{equation}
		H_\ell=a_{-(\ell+1)}^- a_{-(\ell+1)}^+-\frac{\ell}{2}-\frac{3}{4}=a_{-\ell}^+ a_{-\ell}^- -\frac{\ell}{2}+\frac{1}{4},
	\end{equation}
	for the intertwinings
	\begin{equation}
		H_{\ell-1} a_{-\ell}^- = a_{-\ell}^-\left(H_{\ell}-\frac{1}{2}\right),\qquad
		H_{\ell} a_{-\ell}^+ = a_{-\ell}^+\left(H_{\ell-1}+\frac{1}{2}\right),
	\end{equation}
	and for the eigenvalue equations
	\begin{equation}
		H_{\ell-1} (a_{-\ell}^- \psi_{n\ell}) = \left(E_{n\ell}- \frac{1}{2}\right)(a_{-\ell}^-\psi_{n\ell}),\qquad 
		H_{\ell+1} (a_{-(\ell+1)}^+ \psi_{n\ell}) = \left(E_{n\ell}+\frac{1}{2}\right)(a_{-(\ell+1)}^+\psi_{n\ell}).
	\end{equation}
	Neither $a_\ell^\pm$ nor $a_{-\ell}^\pm$  are ladder operators, but through them we can build second-order ones, as can be seen in Figure~\ref{diaRO}, where we show the cases with $\ell={0,1,2,3}$, i.e., for $\ell\in\mathbb{Z}$ even though $\ell$ could be more general, as we stated in Equation \eqref{Hell}. Now, let us take $b_\ell^\pm$ as
	\begin{equation}
		b_\ell^- = a_{-(\ell+1)}^-a_{\ell+1}^-=a_\ell^-a_{-\ell}^-,\qquad
		b_\ell^+ = a_{\ell+1}^+a_{-(\ell+1)}^+=a_{-\ell}^+a_{\ell}^+.
	\end{equation}
	Then, it is easily shown that $H_\ell b_\ell^- = b_\ell^-(H_\ell-1)$ and $H_\ell b_\ell^+ = b_\ell^+(H_\ell + 1)$, i.e., the following commutators are obeyed 
	\begin{equation}
		[H_\ell,b_\ell^{\pm}]=\pm b_\ell^\pm,
	\end{equation}
	which means that $b_\ell^\pm$ are ladder operators with explicit form given by
	\begin{equation}
		b_\ell^\pm=\frac{1}{2}\left(\frac{\text{d}^2}{\text{d}x^2}\mp x\frac{\text{d}}{\text{d}x}+\frac{x^2}{4}-\frac{\ell(\ell+1)}{x^2}\mp\frac{1}{2}\right).
	\end{equation}
	
	It is worth to notice that in the standard formalism of SUSY QM, for $\ell \geq 0$ the replacement $\ell\rightarrow -(\ell+1)$ transforms the superpotential that leads to an unbroken SUSY into a different one, for which SUSY is broken (see \cite{junker}, Section 3.3). Concerning solutions of the Schr\"odinger equation, this phenomenon just means that the transformation maps the ground state eigenfunction of $H_\ell$ into a Schr\"odinger solution for a different factorization energy which does not satisfy anymore the boundary condition at $x=0$ (see below.)
	
	Now, we can obtain the eigenstates of $H_\ell$ departing from the {\it ground state} $\psi_{0\ell}$, an eigenstate of $H_\ell$ such that $b_\ell^- \psi_{0\ell}=0$. In this system there are two such formal eigenstates that satisfy $a_{\ell+1}^-\psi_{\varepsilon_1}= a_{-\ell}^-\psi_{\varepsilon_2}=0$. By solving these first-order differential equations, we obtain:
	\begin{subequations}
		\begin{alignat}{3}
			\psi_{\varepsilon_1} &\propto x^{\ell+1}\exp(-x^2/4), & \qquad \varepsilon_1 & =\frac{\ell}{2}+\frac{3}{4}\equiv E_{0\ell},\label{solsRO1}\\
			\psi_{\varepsilon_2} &\propto x^{-\ell}\exp(-x^2/4), & \qquad \varepsilon_2 & = -\frac{\ell}{2}+\frac{1}{4}= -E_{0\ell}+1.
		\end{alignat}\label{solsRO}
	\end{subequations}
	\hspace{-1mm}As we mentioned earlier, the transformation $\ell\rightarrow -(\ell +1)$ maps $\psi_{\varepsilon_1}\rightarrow\psi_{\varepsilon_2}$ and vice versa. However, $\psi_{\varepsilon_1}$ satisfies the boundary condition \eqref{eigenfun} at $x=0$ for $\ell\geq 0$, but $\psi_{\varepsilon_2}$ does not. From each one of them we can start a ladder of solutions of the Schr\"odinger equation through the operator $b_\ell^+$, but only one of them is physical.
	
\begin{figure}
		\centering
		\includegraphics[scale=0.43]{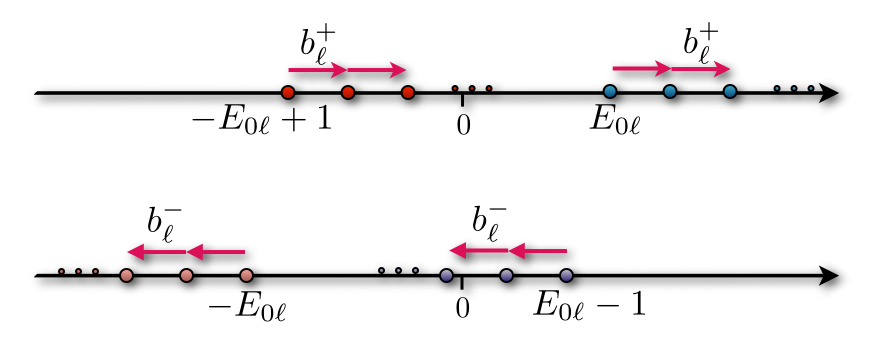}
		\caption{\small{Spectrum of the radial oscillator Hamiltonian $H_\ell$ for $\ell \geq 0$. In the upper part, the blue circles represent the physical solutions starting from $E_{0\ell}$ and the red circles represent the formal ones departing from $-E_{0\ell}+1$, and we go up from these two ladders by $b_\ell^+$. In the lower part, the purple and orange circles represent formal solutions starting from $E_{0\ell}-1$ and $-E_{0\ell}$, and we go down from them by $b_\ell^-$.}}\label{speRO}
	\end{figure}

	We can also obtain two formal eigenfunctions of $H_\ell$ which are simultaneously solutions of the second-order equation $b_\ell^+ \psi_{0\ell}=0$. They satisfy $a_\ell^+\psi_{\varepsilon_3}=a_{-(\ell+1)}^+\psi_{\varepsilon_4}=0$, leading to:
	\begin{subequations}
		\begin{alignat}{3}
			\psi_{\varepsilon_3} &\propto x^{\ell+1}\exp(x^2/4), & \qquad \varepsilon_3 & =\frac{\ell}{2}-\frac{1}{4}= E_{0\ell}-1,\\
			\psi_{\varepsilon_4} &\propto x^{-\ell}\exp(x^2/4), & \qquad \varepsilon_4 & = -\frac{\ell}{2}-\frac{3}{4}= -E_{0\ell},
		\end{alignat}\label{solsRO2}
	\end{subequations}
	\hspace{-1mm}which will be useful later on in our treatment. We can also create from them a ladder of solutions using the operator $b_\ell^-$. Note that for $\ell \geq 0$, of those four solutions, only the first one \eqref{solsRO1} fulfills the boundary conditions and therefore leads to a ladder of ({\it physical}) eigenfunctions of $H_\ell$; the other three formal eigenfunctions of $H_\ell$ supply only mathematical solutions, but they still lead to solutions of PV equation, as we will show in Section \ref{solsPV}. In particular, for $\ell=0$, $\psi_{\varepsilon_2}$ neither diverges nor vanishes at $x=0$, i.e., it does not satisfy the boundary conditions to be an eigenfunction of $H_\ell$. The spectrum of the radial oscillator for $\ell\geq 0$ is therefore
	\begin{equation}
		\text{Sp}(H_\ell)=\{E_{n\ell}=n+\frac{\ell}{2}+\frac{3}{4}, n=0,1,\dots \}.
	\end{equation}
	A diagram of this spectrum can be seen in Figure~\ref{speRO}, where we represent both the physical and formal solutions generated from the extremal states of Equations~\eqref{solsRO} and \eqref{solsRO2}. Moreover, an analogue of the number operator can be defined for the radial oscillator as
	\begin{equation}
		b_\ell^+b_\ell^-=(H_\ell-\varepsilon_1)(H_\ell-\varepsilon_2)=\left(H_\ell-E_{0\ell}\right)\left(H_\ell+E_{0\ell}-1\right),
	\end{equation}
	which is a polynomial of second degree in $H_\ell$, i.e., the radial oscillator is ruled by a first-order PHA.
	
	On the other hand, for $-1/2 < \ell < 0 $ both extremal states of Equation \eqref{solsRO} satisfy the required boundary conditions. Thus, through $b_\ell^+$ they give place to two independent ladder of eigenfunctions of $H_\ell$. However, the scalar product of the states \eqref{solsRO} is not zero, which means that each ladder belongs to a different self-adjoint extension of $H_\ell$ \cite{Wolf81}. Only the first of them, given by Equation \eqref{solsRO1}, leads to a good supersymmetry for the pair $H_\ell, H_{\ell+1}$. Nonetheless, despite this indeterminacy, both extremal states will be quite useful for generating formal solutions to the PV equation in the interval $-1/2 < \ell < 0 $. Moreover, for $\ell=-1/2$ both of these solutions are degenerated so we do not have two different self-adjoint extensions and we can also use our treatment to generate PV solutions there.
	
	Now, in order to implement the SUSY transformations, we employ the general solution of the stationary Schr\"odinger equation for any factorization energy $\epsilon$, which is given by (provided that $\ell$ is not a half-odd number) \cite{CFNN04,JR98,Car01}
	\begin{subequations}
		\begin{align}
			u(x)=\, & x^{-\ell}\text{e}^{-x^2/4}\left[{}_1F_1\left(\frac{1-2\ell-4\epsilon}{4},\frac{1-2\ell}{2};\frac{x^2}{2}\right)\right. \nonumber\\
			& + \left. \nu \frac{\Gamma\left(\frac{3+2\ell-4\epsilon}{4}\right)}{\Gamma\left(\frac{3+2\ell}{2}\right)}\left(\frac{x^2}{2}\right)^{\ell+1/2}{}_1F_1\left(\frac{3+2\ell-4\epsilon}{4},\frac{3+2\ell}{2};\frac{x^2}{2}\right)\right]\label{solRO}\\
			=\, & x^{-\ell}\text{e}^{x^2/4}\left[{}_1F_1\left(\frac{1-2\ell+4\epsilon}{4},\frac{1-2\ell}{2};-\frac{x^2}{2}\right)\right. \nonumber\\
			& + \left. \nu \frac{\Gamma\left(\frac{3+2\ell-4\epsilon}{4}\right)}{\Gamma\left(\frac{3+2\ell}{2}\right)}\left(\frac{x^2}{2}\right)^{\ell+1/2}{}_1F_1\left(\frac{3+2\ell+4\epsilon}{4},\frac{3+2\ell}{2};-\frac{x^2}{2}\right)\right].\label{solROb}
		\end{align}
	\end{subequations}
	
	From these two equivalent expressions of the general solution, we can easily obtain the form of the functions resulting from Equations \eqref{solsRO} and \eqref{solsRO2} that appear represented in the diagram in Figure \ref{speRO}. They are expressed in terms of associated Laguerre polynomials $L_n^{\alpha}$ as
	\begin{subequations}
		\begin{alignat}{3}
			\psi_{1n}(x)&=x^{\ell+1}\text{e}^{-x^2/4}L_n^{\ell+1/2}\left(x^2/2\right),& \qquad E_{1n}&=E_{0\ell}+n,\\
			\psi_{2n}(x)&=x^{-\ell}\text{e}^{-x^2/4}L_n^{-\ell-1/2}\left(x^2/2\right),& \qquad E_{2n}&=-E_{0\ell}+1+n,\\
			\psi_{3n}(x)&=x^{\ell+1}\text{e}^{x^2/4}L_n^{\ell+1/2}\left(-x^2/2\right),& \qquad E_{3n}&=E_{0\ell}-1-n,\\
			\psi_{4n}(x)&=x^{-\ell}\text{e}^{x^2/4}L_n^{-\ell-1/2}\left(-x^2/2\right),& \qquad E_{4n}&=-E_{0\ell}-n,
		\end{alignat}
	\end{subequations}
	where we must remember that $\psi_{1n}$ are the physical solutions for $\ell\geq 0$, while the other ones are mathematical.
	
	Let us take the solution \eqref{solRO} to make a first-order SUSY transformation. Thus, the SUSY partner potential of the radial oscillator becomes
	\begin{equation}
		V_1(x)=\frac{x^2}{8}+\frac{\ell(\ell+1)}{2x^2}- [\ln u(x)]''.
	\end{equation}
	The conditions that must be fulfilled to produce a non-singular transformation are (see \cite{CFNN04}):
	\begin{equation}
		x>0, \qquad \epsilon<E_{0\ell}, \qquad \nu\geq -\frac{\Gamma\left(\frac{1-2\ell}{2}\right)}{\Gamma\left(\frac{1-2\ell-4\epsilon}{4}\right)},\label{condRO}
	\end{equation}
	and the spectrum of $H_1$ becomes Sp$(H_1)=\{\epsilon,E_{0\ell},E_{1\ell},\dots \}$.
	
\begin{figure}\centering
		\includegraphics[scale=0.335]{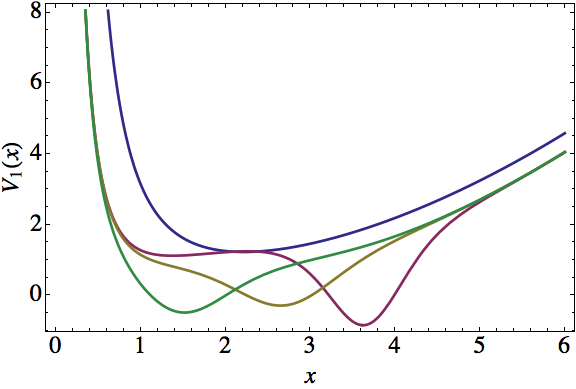}
		\includegraphics[scale=0.35]{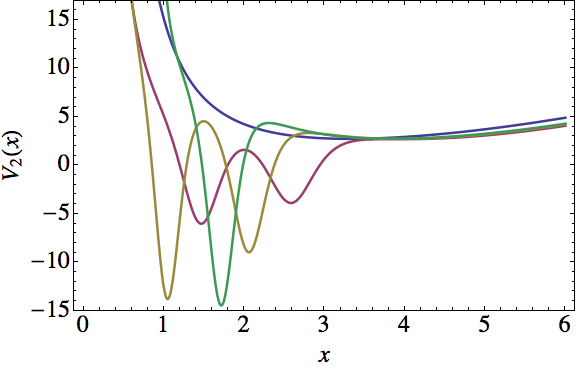}
		\caption{\small{Radial oscillator potential (blue) and its SUSY partners for $k=1$, $\ell=2$, $\epsilon=1/2$, and $\nu=\{-$0.59 (magenta), $-0.4$ (yellow), 1 (green)$\}$ (left plot). The right plot corresponds to $k=2$, $\ell=5$, $\nu_1=1$, and $\epsilon_1=\{$0 (magenta), $-2$ (yellow), $-4$ (green)$\}$ with $u_2 = b_\ell^- u_1$ and $\epsilon_2 = \epsilon_1-1$.}}\label{potRO}
	\end{figure}

	Let us perform a $k$-th order SUSY transformation through $B^\pm_k$ by taking $k$ appropriate solutions $\{u_k,\dots,u_1\}$ in the form given in Equation~\eqref{solRO}, for $k$ factorization energies such that $\epsilon_k<\epsilon_{k-1}<\cdots<\epsilon_1<E_{0\ell}$. The SUSY partner potential of the radial oscillator becomes now:
	\begin{equation}
		V_k(x)=\frac{x^2}{8}+\frac{\ell(\ell+1)}{2x^2}-\{\ln[W(u_1,\dots , u_k)]\}'',
	\end{equation}
	while its spectrum is given by $\text{Sp}(H_k)=\{\epsilon_k,\dots,\epsilon_1,E_{0\ell},E_{1\ell},\dots \}$. In Figure~\ref{potRO} we show examples of first (left) and second-order (right) SUSY partner potentials of the radial oscillator. 
	
	We can define now a natural pair of ladder operators $L_k^{\pm}$ for $H_k$ as 
	\begin{equation}
		L_k^\pm=B_k^+b_\ell^\pm B_k^- ,
	\end{equation}
	that are of $(2k+2)$th-order and fulfill 
	\begin{equation}
		[H_k,L_k^\pm]=\pm L_k^\pm .
	\end{equation}
	From the intertwining relations we can obtain the analogue of the number operator for the $k$-th order SUSY partners of the radial oscillator as
	\begin{equation}
		N(H_k)=L_k^+L_k^-=\prod_{j=1}^{2k+2}(H_k-\varepsilon_j)
		=\left(H_k- E_{0\ell}\right)\left(H_k+E_{0\ell}-1\right)\prod_{j=1}^{k}(H_k-\epsilon_j)(H_k-\epsilon_j-1).
	\end{equation}
	The roots of this polynomial suggest the following: since both $\epsilon_j$ and $\epsilon_j+1$ are in the set, there is a finite ladder  starting and ending at $\epsilon_j$. As the index $j$ runs from $1$ to $k$, this means that Sp($H_k$) contains $k$ one-step ladders. Note that $E_{0\ell}$ and $-E_{0\ell}+1$ are also roots of $N(H_k)$, then Sp($H_k$) could have in principle two infinite ladders, but just the one starting from $E_{0\ell}$ is physical. We conclude that $\{H_k, L_k^-, L_k^+\}$ generates a $(2k+1)$th-order polynomial Heisenberg algebra.
	
	In this article we will work with the simple condition $\epsilon_j \leq E_0$, although it is known that in second-order SUSY QM we can use two transformation functions with $\epsilon_1,\,\epsilon_2$ between two neighbor energy-levels of $H_\ell$ to produce non-singular transformations (see e.g., \cite{FF05}).
	
	\section{Reduction theorem for the SUSY generated Hamiltonians $H_k$}\label{sectmaRO}
	
	In order to implement the prescription pointed out at the end of Section~\ref{topha} to produce solutions to the PV equation, first of all we need to identify systems ruled by a third-order PHA, generated by fourth-order ladder operators. Note that for $k=1$ the operator set $\{H_k, L_k^-, L_k^+\}$ of the previous section generates a third-order PHA, i.e., the first-order SUSY partners of the radial oscillator can be used directly for generating solutions to the PV equation. However, for $k>1$ the set $\{H_k, L_k^-, L_k^+\}$ generates a PHA of order greater than three. Would it be possible to identify a subfamily of the $k$-th order SUSY partners of the radial oscillator Hamiltonian which, in addition of having the natural $(2k+2)$th-order ladder operators $L_k^{\pm}$ would have fourth-order ones? If so, we could generate additional solutions to the PV equation. The answer to this question turns out to be positive, and the conditions required to produce such a reduction process are contained in the following theorem \cite{Ber13}.
	
	\medskip
	
	\noindent {\bf Theorem.} Let $H_k$ be the $k$th-order SUSY partner of the radial oscillator Hamiltonian $H_0$ generated by $k$ Schr\"odinger seed solutions. These solutions $u_i$ are connected by the annihilation operator of the radial oscillator $b_\ell^-$ as
	\begin{equation}
		u_i = (b_\ell^{-})^{i-1} u_1, \qquad \epsilon_i = \epsilon_1 - (i-1), \qquad i=1,\dots,k, \label{restrRO}
	\end{equation}
	where $u_1(x)$ is a Schr\"odinger solution without zeroes, given by Equation~\eqref{solRO} for $\epsilon_1 < E_{0\ell}=\ell/2+3/4$ and
	\begin{equation}
		\nu_1\geq -\frac{\Gamma\left(\frac{1-2\ell}{2}\right)}{\Gamma\left(\frac{1-2\ell-4\epsilon_1}{4}\right)}.
	\end{equation}
	Therefore, the natural $(2k+2)$th-order ladder operator $L_k^+ = B_k^{+} b_\ell^{+} B_k^{-}$ of $H_k$ turn out to be factorized in the form
	\begin{equation}
		L_k^+ = P_{k-1}(H_k) l_k^+,\label{hipoRO}
	\end{equation}
	where $P_{k-1}(H_k) = (H_k - \epsilon_1)\dots(H_k - \epsilon_{k-1})$ is a polynomial of degree $k-1$ in $H_k$ and $l_k^+$ is a fourth-order differential ladder operator,
	\begin{equation}
		[H_k,l_k^+] = l_k^+, \label{conmHlRO}
	\end{equation}
	such that
	\begin{equation} \label{annumk3RO}
		l_k^+ l_k^- =\left(H_k -E_{0\ell} \right)\left(H_k + E_{0\ell} -1 \right)(H_k - \epsilon_k)(H_k - \epsilon_1-1).
	\end{equation}
	\medskip
	
	\noindent {\bf Proof (by induction).}
	
	For $k=1$ the result is straightforward
	\begin{equation}
		L_1^+ = P_0(H_1)l_1^+ , \qquad P_0(H_1) = 1.
	\end{equation}
	
	Let us suppose now that the theorem is valid for a given $k$ (induction hypothesis) and then we are going to show that it is also valid for $k+1$, i.e., we assume that $L_{k}^{+}=P_{k-1}(H_k)l_{k}^{+}$ and we will proof that $L_{k+1}^{+}=P_{k}(H_{k+1})l_{k+1}^{+}$.
	
	From the intertwining technique it is clear that we can go from $H_k$ to $H_{k+1}$ and vice versa through a first-order SUSY transformation
	\begin{equation}
		H_{k+1} A_{k+1}^+ = A_{k+1}^+ H_k, \qquad H_kA_{k+1}^{-}=A_{k+1}^{-}H_{k+1}.
	\end{equation}
	Moreover, it is straightforward to show that
	\begin{equation}
		L_{k+1}^+ = A_{k+1}^+ L_{k}^+ A_{k+1}^- .\label{LentreRO}
	\end{equation}
	
	From the induction hypothesis one obtains
	\begin{equation}
		L_{k+1}^{+}  = A_{k+1}^{+}P_{k-1}(H_k)l_{k}^{+}A_{k+1}^{-}
		= P_{k-1}(H_{k+1})\underbrace{A_{k+1}^{+}l_{k}^{+}A_{k+1}^{-}}_{\widetilde{l}_{k+1}^{+}},\label{plRO}
	\end{equation}
	where
	\begin{equation}
		\widetilde{l}_{k+1}^{+}\equiv A_{k+1}^{+}l_{k}^{+}A_{k+1}^{-},\label{l52RO}
	\end{equation}
	is a sixth-order differential ladder operator for $H_{k+1}$. A direct calculation leads to
	\begin{equation}
		\widetilde{l}_{k+1}^+ \widetilde{l}_{k+1}^- = (H_{k+1} - \epsilon_k)^2 \left(H_{k+1} -E_{0\ell} \right)\left(H_{k+1} + E_{0\ell} -1 \right)(H_{k+1} - \epsilon_{k+1})(H_{k+1} - \epsilon_1-1).
	\end{equation}
	Note that the last four factors in the right hand side of this equation are precisely what would be obtained from the product $l_{k+1}^+ l_{k+1}^-$ of the fourth-order ladder operators of $H_{k+1}$. Thus, it is concluded that
	\begin{equation}
		\widetilde{l}_{k+1}^+ = q(H_{k+1}) l_{k+1}^+,
	\end{equation}
	where $q(H_{k+1})$ is a polynomial in $H_{k+1}$. By remembering that $\widetilde{l}_{k+1}^+, l_{k+1}^+$, and $H_{k+1}$ are differential operators of sixth, fourth, and second order respectively, one can conclude that $q(H_{k+1})$ is linear in $H_{k+1}$. As we already know that $\epsilon_k$ is a root of $q(H_{k+1})$ we get
	\begin{equation}
		\widetilde{l}_{k+1}^+ = (H_{k+1} - \epsilon_k) l_{k+1}^+ .\label{N52RO}
	\end{equation}
	By substituting this result in Equation~\eqref{plRO} we finally obtain
	\begin{equation}
		L_{k+1}^+ = P_{k-1}(H_{k+1})(H_{k+1} - \epsilon_k) l_{k+1}^+ = P_{k}(H_{k+1})l_{k+1}^+,
	\end{equation}
	which concludes our proof. \hfill $\square$
	
	\subsection{Properties of the operators $l_k^{\pm}$}\label{secope5}
	
	First of all, let us remember that the natural ladder operators $L_k^\pm$ connect just the eigenstates associated to the initial part of the spectrum $E_{n\ell}$, but annihilate all the eigenstates for the newly created levels at $\epsilon_i$.
	
\begin{figure}\centering
		\includegraphics[scale=0.35]{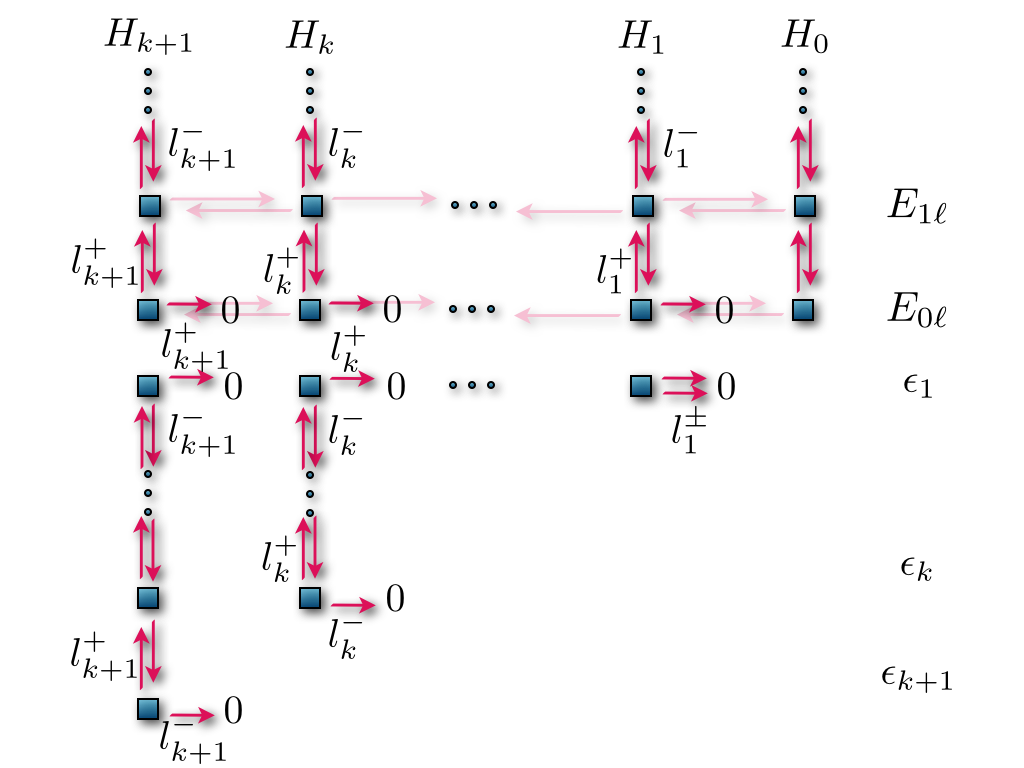}
		\caption{\small{Action of the operators $l_{k}^{\pm}$ over the eigenstates of the SUSY Hamiltonians $H_k$. Note that $l_1^{\pm}=L_1^{\pm}$, and one can see that the operator $l_{k}^{-}$ always annihilates the eigenstates associated with $E_{0\ell}$ and $\epsilon_k$, while $l_{k}^{+}$ annihilates the one associated with $\epsilon_1$.}}\label{fig.tma2RO}
	\end{figure}

	On the other hand, the reduced ladder operators $l_k^\pm$ do actually allow the displacement between the eigenstates of the finite ladder. Moreover, in the physical sector the operator $l_{k}^{-}$ annihilates only the eigenstates associated with $E_{0\ell}$ (the initial ground state energy) and with $\epsilon_k$ (the new ground state level), while $l_{k}^{+}$ annihilates only the new eigenstate for the energy $\epsilon_1$. A diagram representing the action of the fourth-order ladder operators $l_k^{\pm}$ on the eigenstates of the SUSY Hamiltonians $H_k$ is shown in Figure~\ref{fig.tma2RO}.
	
	\subsubsection{Relation with $A_{k+1}^{\pm}$.}
\begin{figure}\centering
		\includegraphics[scale=0.35]{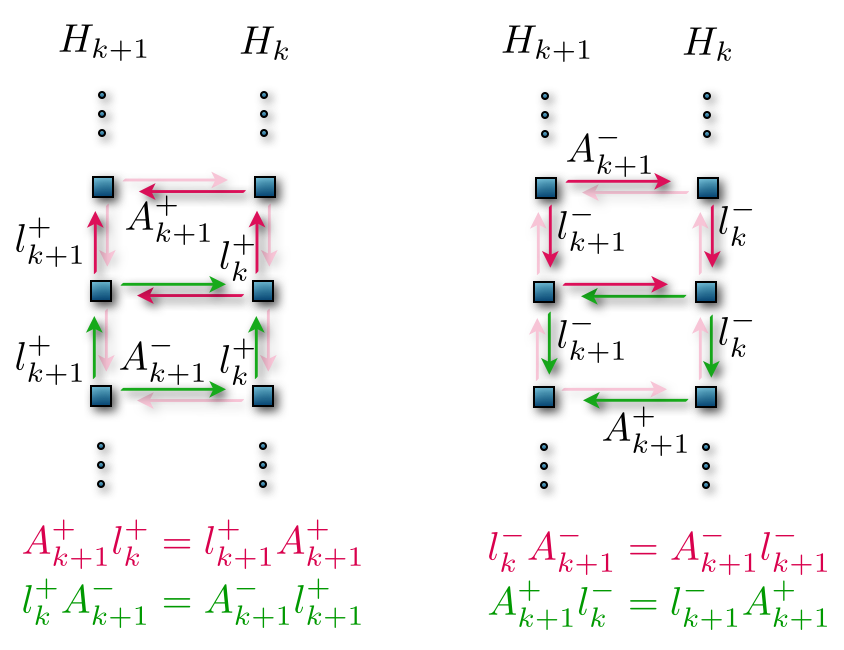}
		\caption{\small{Action of the operators from Equations~\eqref{finalAlRO} over the eigenstates of the SUSY Hamiltonians $H_k$ and $H_{k+1}$. We can see that these relations allow the displacement to any level in the spectrum, including the new eigenstates.}}\label{fig.tma4RO}
	\end{figure}

	There are interesting relations involving the reduced ladder operators $l_k^\pm$ and the first-order intertwining ones $A_{k+1}^{\pm}$, namely,
	\begin{subequations}
		\begin{align}
			A_{k+1}^{+}l_{k}^{\pm} &= l_{k+1}^{\pm}A_{k+1}^{+},\\
			l_{k}^{\pm}A_{k+1}^{-} &= A_{k+1}^{-}l_{k+1}^{\pm}.
		\end{align}\label{finalAlRO}
	\end{subequations}
	\hspace{-1mm}Note that these four relations are general, i.e., they can be applied to any eigenstate in the physical ladders, including those which are annihilated by operators involved in such a relations. A full diagram can be seen in Figure~\ref{fig.tma4RO}.
	
	\subsubsection{Analogue of the number operator $l_{k}^{+}l_k^{-}$.}
	Let us recall that the analogue of the number operator $b_\ell^{+}b_\ell^{-}$ acts on the eigenstates $|\psi_{n}^{(0)}\rangle$ of the radial oscillator Hamiltonian $H_0$ as a multiplication by a second degree polynomial in $n$, i.e.,
	\begin{equation}
		b_\ell^{+}b_\ell^{-}|\psi_{n}^{(0)}\rangle = n(n+2E_{0\ell}-1) |\psi_{n}^{(0)}\rangle.
	\end{equation}
	On the other hand, for its SUSY partner Hamiltonians $H_k$ the action of the number operator, built through the natural $(2k+2)$th-order ladder operators ($L_k^{+}L_k^{-}$), onto the eigenstates of $H_k$ becomes
	\begin{subequations}
		\begin{align}
			L_k^{+}L_k^{-}|\psi_{n}^{(k)}\rangle &=\left[ n(n+2E_{0\ell}-1) \prod_{i=1}^{k}\left(n+E_{0\ell}-\epsilon_i \right)\left(n+E_{0\ell}-\epsilon_i-1\right)\right] |\psi_{n}^{(k)}\rangle ,\\
			L_k^{+}L_k^{-}|\psi_{\epsilon_j}^{(k)}\rangle &= 0.
		\end{align}
	\end{subequations}
	Now, for the fourth-order ladder operators $l_k^{\pm}$ of $H_k$, the analogue of the number operator $l_k^{+}l_k^{-}$ acts onto the eigenstates of $H_k$ as follows
	\begin{subequations}
		\begin{align}
			l_k^{+}l_k^{-} | \psi_{n}^{(k)}\rangle &=n(n+2E_{0\ell}-1)\left(n+E_{0\ell}-\epsilon_1 -1\right)\left(n+E_{0\ell}-\epsilon_k\right) | \psi_{n}^{(k)}\rangle , \\
			l_k^{+}l_k^{-} | \psi_{\epsilon_i}^{(k)}\rangle &= (\epsilon_i-E_{0\ell})(\epsilon_i+E_{0\ell}-1)(\epsilon_i -\epsilon_1 -1)(\epsilon_i-\epsilon_k) | \psi_{\epsilon_i}^{(k)}\rangle .
		\end{align}
	\end{subequations}
	We should note that the only physical eigenstates which are annihilated by $l_k^{+}l_k^{-}$ are those associated with the old ground state energy $E_{0\ell}$ and the new lower level $\epsilon_k$.
	
	\subsection{Consequences of the theorem}\label{secnuevasgRO}
	At the beginning of this section we have proven a theorem that establishes the conditions under which the following factorization is fulfilled
	\begin{equation}
		L_{k}^{+}=P_{k-1}(H_k)l_{k}^{+},\label{facttmaRO}
	\end{equation}
	i.e., for the natural ladder operators $L_k^\pm$ to be expressed as products of a polynomial of degree $k-1$ in $H_k$ times the fourth-order ladder operators $l_k^\pm$. This means that the $(2k+1)$th-order PHA, obtained through a SUSY transformation involving $k$ connected seed solutions as specified in the theorem, with $\epsilon_i=\epsilon_1-(i-1),\ i=1,\dots ,k$, can be \emph{reduced} to a third-order PHA with fourth-order ladder operators \cite{AS05}.
	
	We recall from Section~\ref{topha} that these algebras are closely related to the PV equation. This means that when we reduce the higher-order algebras, we open the possibility of obtaining new solutions of the PV equation, similar to what happens for second-order PHA and PIV equation \cite{BF11a,Ber13,BF11b,Ber12,BF12,BF13a}. In the following sections we will introduce our method to obtain solutions of the PV equation, and then we will see that for some cases they can be written in terms of well known special functions; this will lead us to classify them into several solution hierarchies.
	
	\section{Solutions to PV equation through SUSY QM}\label{solsPV}
	
	As it was pointed out previously, in order to generate solutions to the PV equation we have to find systems ruled by a third-order PHA, having fourth-order ladder operators. In fact, the key point of the technique is to identify the four extremal states (either physical or mathematical) of the system, as well as their associated energies. Once identified, we number these states arbitrarily (from first to fourth) and use Equations~\eqref{gdex} and \eqref{wdez} to calculate the PV solution and Equation~\eqref{paraPV} to determine the parameters. Since both, the PV solution and the parameters of the equation, are symmetric under the exchanges $\varepsilon_1 \leftrightarrow \varepsilon_2$ and $\varepsilon_3 \leftrightarrow \varepsilon_4$, then from the $4!=24$ possible permutations of the four indexes we will get just six different solutions to the PV equation, some of which could have singularities.
	
	Let us apply the technique next to the radial oscillator potential, then to its first-order SUSY partners and, finally, to the subfamily of $k$-th order SUSY partners which also have the fourth order differential ladder operators $l_k^\pm$.
	
	\subsection{Radial oscillator}
	
	We have seen previously that the radial oscillator Hamiltonian has second-order differential ladder operators $b_\ell^\pm$ leading to a first-order PHA. We can also construct several pairs of fourth-order differential ladder operators giving place to a third-order PHA, for example,
	\begin{equation}
		L_4^\pm = a_{\ell+1}^+ b_{\ell+1}^\pm a_{\ell+1}^-,
	\end{equation}
	although three additional operators can be inferred from Figure \ref{diaRO}.
	The analogue to the number operator becomes in this case:
	\begin{equation}
		L_4^+ L_4^- =
		\left(H_\ell + E_{0\ell} - 1 \right)\left(H_\ell - E_{0\ell}\right) \left(H_\ell - E_{1\ell}\right)^2.
	\end{equation}
	Its roots suggest naturally three extremal states: two of them are those associated to the first-order PHA, and another one is the first excited state $\psi_{1\ell}$ of $H_\ell$ with eigenvalue $E_{1\ell}$. The last one is chosen as the mathematical eigenstate of  $H_\ell$ associated to the eigenvalue $E_{1\ell}$, denoted as $\psi_{1\ell}^\perp$, such that $W(\psi_{1\ell},\psi_{1\ell}^\perp)=1$. Now, we number them as follows:
	\begin{subequations}
		\begin{alignat}{3}
			\psi_{\varepsilon_1} &\propto x^{\ell+1}\exp(-x^2/4), & \qquad \varepsilon_1 & = E_{0\ell},\\
			\psi_{\varepsilon_2} &\propto x^{-\ell}\exp(-x^2/4), & \qquad \varepsilon_2 & =  -E_{0\ell}+1, \\
			\psi_{\varepsilon_3}& = \psi_{1\ell}, &  \qquad  \varepsilon_3 & = E_{1\ell}, \\
			\psi_{\varepsilon_4}& = \psi_{1\ell}^\perp, & \qquad  \varepsilon_4 & = E_{1\ell}.
		\end{alignat}\label{extremalRO1}
	\end{subequations}
	\hspace{-1mm}Using these expressions and Equations~\eqref{paraPV}, \eqref{gdex} and \eqref{wdez}, it is straightforward to obtain the PV solution with parameters:
	\begin{equation}
		a= \frac{(2\ell + 1)^2}{8}, \qquad b = 0, \qquad c=-\frac{2\ell + 7}{4}, \qquad d = - \frac18.
	\end{equation}
	In Equation \eqref{paraPV} the four parameters of the PV equation are expressed in terms of the four extremal states energies, but we also have symmetry in the exchanges $\varepsilon_1 \leftrightarrow \varepsilon_2$ and $\varepsilon_3 \leftrightarrow \varepsilon_4$. Thus from the $4!=24$ possible permutations of the four indexes we have just six different solutions to PV equation. Now, by exploring all possible permutations of the indexes which lead to different PV solutions, the six results contained in Table~\ref{table0} are obtained.
	
	\begin{table}
		\begin{center}
			\begin{tabular}{ccccc}
				\hline
				Order & $8a$&$8b$&$4c$&$w(z)-1$\\
				\hline
				$1234$ &		$(2\ell+1)^2$ 	&	$0$				&		$-2\ell-7$  & 	-1	\\
				$1324$ &		$4$ 				&	$-(2\ell+3)^2$	&		$2\ell-1$ 	&
				$\frac{z \Gamma (\ell +1/2,0,-z/2)-2^{-\ell+1/2}\text{e}^{z/2}(-z)^{\ell+1/2}}{(2\ell-1)\Gamma (\ell +1/2,0,z/2)}$	\\
				$1423$ &		$4$ 				&	$-(2\ell+3)^2$	&		$2\ell-1$ 	& 	$(2\ell -1)^{-1}z$	\\
				$2314$ &		$(2\ell+3)^2$ 	&	$-4$				&		$-2\ell-3$	& $\frac{(2\ell+z+1)[2^{-1}(2\ell+1)\Gamma(\ell+1/2,-z/2)-\Gamma(\ell+3/2)]+(2\ell+1)(-z/2)^{\ell+1/2}\text{e}^{z/2}}
				{2\Gamma(\ell+3/2)-(2\ell+1)\Gamma(\ell+1/2,-z/2)}$\\
				$2413$ &		$(2\ell+3)^2$ 	&	$-4$				&		$-2\ell-3$	& 	$2^{-1}(1-2\ell-z)$	\\
				$3412$ &		$0$				&	$-(2\ell+1)^2$	&		$2\ell+3$	& 	$\infty$	\\
				\hline
			\end{tabular}
		\end{center}
		\vspace{-3mm}
		\caption{Explicit solutions of the PV equation for the six permutations of the extremal states \eqref{extremalRO1} of the radial oscillator potential.} \label{table0}
	\end{table}
	
	We must remember that, since the radial oscillator is also described by a first-order PHA, then the third-order case must reduce to the original first-order PHA. Indeed, this is what happens since $L_4^+ = b_\ell^+ (H_\ell - E_{0\ell})$ and $L_4^- = (H_\ell - E_{0\ell}) b_\ell^-$. The same can be done for the other three different choices of $L_4^\pm$.
	
	\subsection{First-order SUSY partners of the radial oscillator}
	
	The first-order SUSY partners of the radial oscillator Hamiltonian possess a pair of natural fourth-order differential ladder operators
	\begin{equation}
		L_4^\pm = A_1^+ b_\ell^\pm  A_1^-,
	\end{equation}
	where $A_1^\pm$ are the first-order intertwining operators and $b_\ell^\pm$ are the second-order ladder operators of Section~\ref{SUSYRO}. The operators $L_4^\pm$ give place to a third-order PHA since
	\begin{equation}
		N(H_1) = L_4^+ L_4^- = \left(H_\ell + E_{0\ell} - 1 \right)\left(H_\ell - E_{0\ell}\right) \left(H_\ell - \epsilon \right)\left(H_\ell - \epsilon - 1\right).
	\end{equation}
	The roots of $N(H_1)$ and the SUSY procedure suggest now the following extremal states: two of them are the SUSY transformed extremal states of the radial oscillator, another one is the new ground state created by the SUSY transformation. The last one will be a mathematical eigenstate of $H_1$ associated to $\epsilon +1$. Let us choose now the following ordering:
	\begin{subequations}
		\begin{alignat}{3}
			\psi_{\varepsilon_1}& \propto A_1^+b_\ell^{+}u, & \qquad \varepsilon_1 & = \epsilon + 1,\\
			\psi_{\varepsilon_2}& \propto A_1^+\left[x^{-\ell}\exp\left(-\frac{x^2}{4}\right)\right], & \qquad \varepsilon_2 & =  - E_{0\ell}+1,\\
			\psi_{\varepsilon_3}& \propto \frac{1}{u}, &  \qquad  \varepsilon_3 & = \epsilon,\\
			\psi_{\varepsilon_4}& \propto A_1^+\left[x^{\ell+1} \exp\left(-\frac{x^2}{4}\right)\right], & \qquad  \varepsilon_4 & = E_{0\ell}.
		\end{alignat}\label{extremalRO2}
	\end{subequations}
	
	Using these expressions and Equations~\eqref{paraPV}, \eqref{gdex} and \eqref{wdez}, one can obtain the PV solution with parameters:
	\begin{equation}
		a= \frac{(4\epsilon + 2\ell + 3)^2}{32}, \qquad b = - \frac{(4\epsilon - 2\ell - 3)^2}{32}, \qquad c=-\frac{2\ell + 1}{4}, \qquad d = - \frac{1}{8}.
	\end{equation}
	In addition, we can permute the indexes to obtain different PV solutions. For the first-order SUSY partners of the radial oscillator, although the PV solutions can be quickly calculated with any symbolic software, they are quite large to be written down explicitly; for that reason we have chosen a simple case, using as seed solution $u(x)$ the ground state of the radial oscillator, i.e., $\epsilon=E_{0\ell}$. In this case, of the six possible solutions obtained by permutations, three of them reduce to zero and the other three become fractions of polynomials. The full explicit expressions are given in Table~\ref{table1}.
	
	\begin{table}
		\begin{center}
			\begin{tabular}{ccccc}
				\hline
				Order & $8a$&$8b$&$4c$&$w(z)-1$\\
				\hline
				$1234$ &	$(2\ell+3)^2$ 	&	$0$				&	$-2\ell-1$ 	& 	0	\\
				$1324$ &	$4$ 				&	$-(2\ell+1)^2$	&	$2\ell+1$ 	& 	$0$	\\
				$1423$ &	$4$ 				&	$-(2\ell+1)^2$	&	$2\ell+1$ 	& 	$z(2\ell-z+1)^{-1}$	\\
				$2314$ &	$(2\ell+1)^2$ 	&	$-4$				&	$-2\ell-5$	& $0$\\
				$2413$ &	$(2\ell+1)^2$ 	&	$-4$				&	$-2\ell-5$	& $\frac{z[8 \ell^3-4 \ell^2(z-1)-2\ell (5 z^2+2 z+2)+5 (z-3) z^2]}
				{-8 \ell^3 (z-4)+4 \ell^2 (z^2-3 z+4)+2 \ell (5 z^3+2 z^2-2 z-8)-5 (z-1) z^3}$	\\
				$3412$ &	$0$				&	$-(2\ell+3)^2$	&	$2\ell-3$	& 
				$\frac{z [8 \ell^3+4 \ell^2 - 2\ell (2 + 5 z^2) - 15 z^2]}{16\ell (2 \ell^2+ \ell-1)}$	\\
				\hline
			\end{tabular}
		\end{center}
		\vspace{-3mm}
		\caption{Explicit solutions to PV equation for the six permutations of the extremal states \eqref{extremalRO2} for the first-order SUSY partner of the radial oscillator with $\epsilon=E_{0\ell}$.} \label{table1}
	\end{table}
	
	\subsection{$k$th-order SUSY partners of the radial oscillator}
	
	The $k$th-order SUSY partners of the radial oscillator Hamiltonian, generated by $k$ seed solutions which are connected through the second-order ladder operators as
	\begin{equation}
		u_{i}\propto (b_\ell^-)^{i-1} u_1, \qquad \epsilon_{i}=\epsilon_1-(i-1), \qquad i=1,\dots,k,
	\end{equation}
	have fourth-order differential ladder operators $l_k^\pm$ defining the following analogue of the number operator:
	\begin{equation}
		N(H_k) = l_k^+ l_k^- = \left(H_k + E_{0\ell} - 1 \right)\left(H_k - E_{0\ell}\right) \left(H_k - \epsilon_k \right)\left(H_k - \epsilon_1 - 1\right).
	\end{equation}
	Its roots suggest that two extremal states of $H_k$ are the SUSY transformed extremal states of the radial oscillator, another one is the new ground state created by the SUSY transformation at $\epsilon_k$, and the last one is a formal eigenstate of $H_k$ associated to $\epsilon_1 + 1$. We order them as follows:
	\begin{subequations}
		\begin{alignat}{3}
			\psi_{\varepsilon_1}& \propto B_k^+b_\ell^{+}u_1, & \qquad \varepsilon_1 & = \epsilon_1 + 1,\\
			\psi_{\varepsilon_2}& \propto B_k^+\left[x^{-\ell}\exp\left(-x^2/4\right)\right], & \qquad \varepsilon_2 & =  - E_{0\ell}+1,\\
			\psi_{\varepsilon_3}& \propto \frac{W(u_1,\dots,u_{k-1})}{W(u_1,\dots,u_k)}, &  \qquad  \varepsilon_3 & = \epsilon_k,\\
			\psi_{\varepsilon_4}& \propto B_k^+\left[x^{\ell+1} \exp\left(-x^2/4\right)\right], & \qquad  \varepsilon_4 & = E_{0\ell},\label{RO34}
		\end{alignat}\label{extremalRO3}
	\end{subequations}
	where $B_k^+$ is the $k$th-order intertwining operator. 
	
	Using these expressions and Equations~\eqref{paraPV},\eqref{gdex},\eqref{wdez}, the PV solution and its associated parameters are obtained:
	\begin{equation}
		a= \frac{(4\epsilon_1 + 2\ell + 3)^2}{32}, \qquad b = - \frac{(4\epsilon_1 - 4k - 2\ell + 1)^2}{32}, \qquad c=\frac{2k-2\ell - 3}{4}, \qquad d = - \frac18.
	\end{equation}
	In particular, for second-order SUSY ($k=2$) with $\epsilon_1=E_{1\ell}$ and $u_1$ being the eigenfunction of the first excited state of $H_\ell$, the PV solutions are simple. In fact, by exploring all permutations of the indexes generating different PV solutions, we obtain the six results contained in Table \ref{table2}.
	
	\begin{table}
		\begin{center}
			\begin{tabular}{ccccc}
				\hline
				Order & $8a$				&$8b$					&$4c$				& $w(z)$\\
				\hline
				$1234$ &	$(2\ell+5)^2$ 	&	$0$				&	$-2\ell+1$ 	& $0$\\
				$1324$ &	$16$ 				&	$-(2\ell+1)^2$	&	$2\ell+3$	& $0$\\
				$1423$ &	$16$ 				&	$-(2\ell+1)^2$	&	$2\ell+3$ 	& $\frac{4(z-2\ell-3)}{z^2-2z(2\ell+1)+4\ell^2+8\ell+3}$\\
				$2314$ &	$(2\ell+1)^2$ 	&	$-16$				&	$-2\ell-7$	& $0$\\
				$2413$ &	$(2\ell+1)^2$ 	&	$-16$				&	$-2\ell-7$	& $\frac{(-z+2\ell+3)(2\ell+1)}{z^2-2z(2\ell+1)+4\ell(\ell-2)+3}$\\
				$3412$ &	$0$				&	$-(2\ell+5)^2$	&	$2\ell-5$	& $\infty$\\
				\hline
			\end{tabular}
		\end{center}
		\vspace{-3mm}
		\caption{Explicit solutions to PV equation for the six permutations of the extremal states \eqref{extremalRO3} for the second-order SUSY partners of the radial oscillator with $\epsilon_1=E_{1\ell}$.} \label{table2}
	\end{table}
	
	\section{Real solutions to the PV equation with real parameters}\label{rr}
	
	The non-singular SUSY transformations involving real seed solutions, associated with real factorization energies, have one restriction: the whole finite ladder which is created has to be placed below the ground state energy of $H_\ell$, i.e., $\epsilon_1\leq E_{0\ell}$. This kind of transformations gives place to real solutions to the PV equation with real parameters $a,b,c,d$. Let us analyze next some of those solutions and their associated parameters.
	
	\subsection{First-order SUSY QM}
	
\begin{figure}\centering
		\includegraphics[scale=0.27]{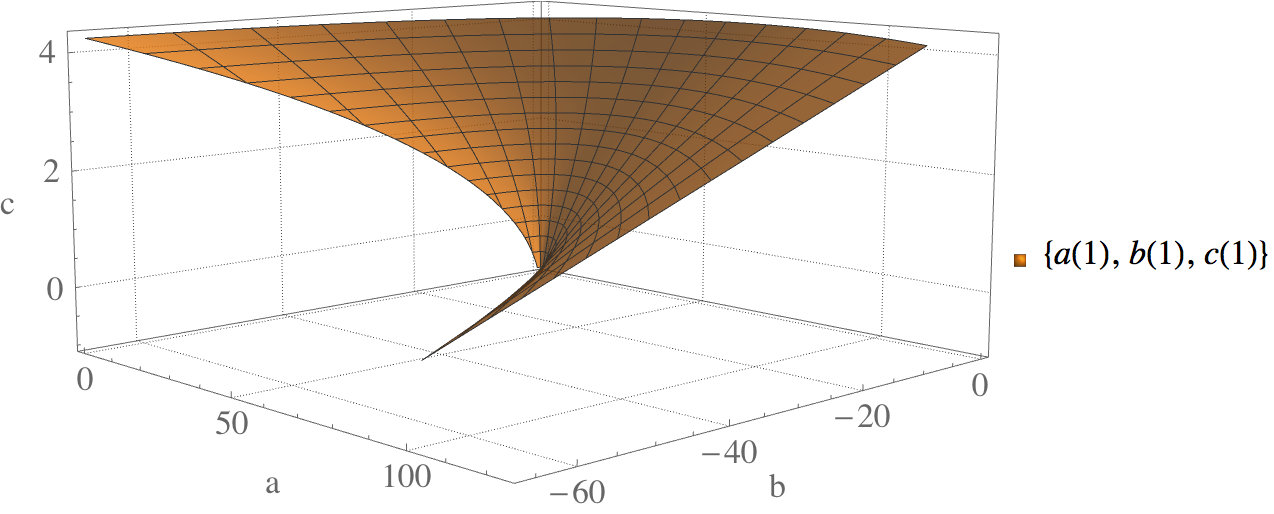}
		\caption{\small{Surface in the parameter space $(a,b,c)$ where there are non-singular real solutions of the PV equation for the ordering 1234 of the extremal states given in Equation \eqref{extremalRO2}. We vary the parameters $\epsilon_1\in (-10,E_{0\ell})$ and $\ell\in (-1/2,10)$. The labels for the legend refer to $k=1$.}}\label{paraRO}
	\end{figure}

	For the first-order SUSY partners of the radial oscillator we are able to connect directly with the PV equation and specific parameters $a,b,c,d\in\mathbb{C}$, in general. In fact, from Equations \eqref{paraPV}, \eqref{eeps} and \eqref{extremalRO2}, we obtain $a,b,c,d$ in terms of one parameter of the initial system ($E_{0\ell}$) and one of the SUSY transformation ($\epsilon_1$) as follows:
	\begin{equation}
		a=\frac{(E_{0\ell}+\epsilon_1)^2}{2},\qquad b=-\frac{(E_{0\ell}-\epsilon_1)^2}{2},\qquad c=\frac{1-2E_{0\ell}}{2},\qquad d=-\frac{1}{8}.
	\end{equation}
	Since $E_{0\ell}=\ell/2+3/4$, in general the four parameters $a,b,c,d$ depend on $\ell\geq-1/2$ and $\epsilon_1\in\mathbb{C}$. In this section we will study the case where the PV parameters and the factorization energy are real, i.e., $a,b,c,d,\epsilon_1\in\mathbb{R}$, while in the next sections we will study the complex case. We must remark that usually in the physical studies of the radial oscillator $\ell$ is the angular momentum index and it is restricted by $\ell\in\mathbb{Z}^+$, but here we employ it just as an aid mechanism for obtaining solutions to the PV equation; that is why we rather use the generalized radial oscillator for $\ell\in\mathbb{R}$ and $\ell\geq -1/2$. In Figure~\ref{paraRO}, we show a parametric plot of the three parameters $a,b,c$ as functions of $\ell$ and $\epsilon_1$ ($d$ does not depend on them). On any point of this surface in the parameter space we can find solutions to the PV equation.
	
	In fact, if we restrict ourselves to real solutions of the PV equation with real parameters and without singularities in $x \geq 0$, we also get the restriction $\epsilon_1\leq E_{0\ell}$. Moreover, for each one of those points we have indeed a one-parameter family of solutions, labeled by the parameter $\nu_1$ from Equation~\eqref{solRO} with the restriction \eqref{condRO}.
	
\begin{figure}\centering
		\includegraphics[scale=0.37]{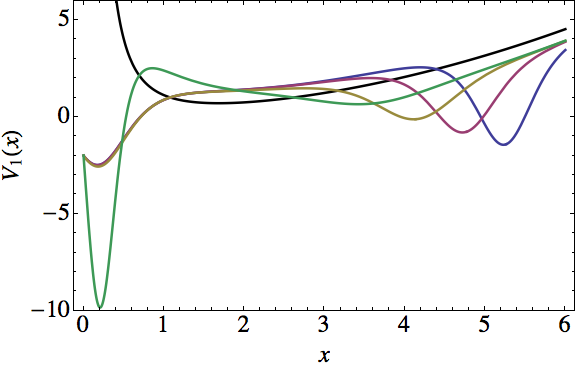}
		\includegraphics[scale=0.37]{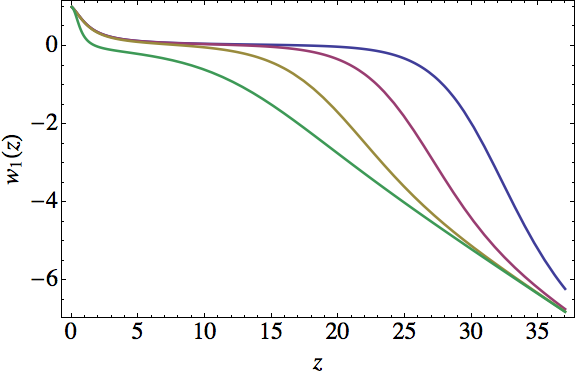}
		\caption{\small{First-order SUSY partner potential (left) $V_1(x)$ of the radial oscillator (black) and the solutions (right) $w_1(z)$ to the PV equation for $\ell=1$, $\epsilon_1=1$, and $\nu_1=\{$0.905 (blue), 0.913 (magenta), 1 (yellow), 10 (green)$\}$.}}\label{pv1}
	\end{figure}

	Then, we obtain the following 1-SUSY partner potential and the function $g_1(x)$, which is related with the solution $w(z)$ of the PV equation:
	\begin{subequations}
		\begin{align}
			V_1(x)&=\frac{x^2}{8}+\frac{\ell(\ell+1)}{2x^2}-[\ln u(x)]'',\\
			g_1(x)&=-\frac{x}{2}-\frac{\ell+1}{x}+[\ln u(x)]',
		\end{align}
	\end{subequations}
	where we have added an index to indicate the order of the SUSY transformation. Since $g_1(x)$ is connected with the solution $w_1(z)$ of the PV equation through
	\begin{equation}
		w_1(z)=1+\frac{z^{1/2}}{g_1(z^{1/2})},
	\end{equation}
	then
	\begin{equation}
		w_1(z)=1+\frac{2zu(z^{1/2})}{2z^{1/2}u'(z^{1/2})-(z+2\ell+2)u(z^{1/2})}.
	\end{equation}
	An illustration of several first-order SUSY partner potentials of the radial oscillator $V_1(x)$ and the corresponding solutions $w_1(z)$ of the PV equation are shown in Figure~\ref{pv1}.
	
	\subsection{$k$th-order SUSY QM}
	With the reduction theorem used to identify the fourth-order ladder operators for the SUSY partners of the radial oscillator of Section~\ref{sectmaRO}, we are able to reduce the ($2k+1$)th-order PHA induced by the natural ladder operators to third-order PHA. Basically, the $k$ transformation functions have to be connected through the annihilation operator $b_\ell^-$ and therefore their energies will be given by $\epsilon_i=\epsilon_1-(i-1)$. This implies that we create a new equidistant ladder with $k$ steps, one step for each first-order SUSY transformation. There is also the restriction on the factorization energy that $\epsilon_1<E_{0\ell}$. 
	
	Once again, we need to identify the extremal states of our system. Since the roots of the polynomial in Equation~\eqref{annumk3RO} are
	$E_{0\ell}, -E_{0\ell} +1, \epsilon_k, \epsilon_1+1$, then two of them are physical extremal states, the ones associated with $E_{0\ell}$ and $\epsilon_k$, a mathematical one coming from the radial oscillator at $-E_{0\ell}+1$, and another mathematical eigenstate at $\epsilon_1+1$ that will make the new ladder to be finite. The four extremal states are thus given by Equation \eqref{extremalRO3}.
	
	From SUSY QM usual theory \cite{Ber13}, we can write $\psi_{\varepsilon_4} $ as
	\begin{equation}
		\psi_{\varepsilon_4} \propto B_k^+ \left[x^{\ell+1} \exp(-x^2/4)\right]\propto \frac{W(u_1,\dots , u_k,x^{\ell+1}\exp(-x^2/4))}{W(u_1,\dots , u_k)}.
	\end{equation}
	Thus, from Equation \eqref{ache} we obtain the auxiliary function $h(x)$ as
	\begin{equation}
		h(x)=\left\{\ln\left[W(\psi_{\varepsilon_3},\psi_{\varepsilon_4})\right]\right\}',
	\end{equation}
	and then from Equation \eqref{gdex} one arrives at
	\begin{equation}
		g(x)=-x-\left\{\ln\left[W(\psi_{\varepsilon_3},\psi_{\varepsilon_4})\right]\right\}'.
	\end{equation}
	Therefore, the $k$th-order SUSY partner potential $V_k(x)$ of the radial oscillator and its corresponding $g_k(x)$ function are
	\begin{subequations}
		\begin{align}
			V_k(x)&=\frac{x^2}{8}+\frac{\ell(\ell+1)}{2x^2}-[\ln W(u_1,\dots , u_k)]'',\\
			g_k(x)&=-x+\frac{2(E_{0\ell}-\epsilon_1+k-1)W(u_1,\dots ,u_{k-1})W(u_1,\dots ,u_{k},x^{\ell+1}\exp(-x^2/4))}{W\left(W(u_1,\dots ,u_{k-1}), W(u_1,\dots ,u_{k},x^{\ell+1}\exp(-x^2/4))\right)}.\label{VgkROb}
		\end{align}\label{VgkRO}
	\end{subequations}
	\hspace{-1mm}Recall that $g_k(x)$ is directly related with $w_k(z)$ through
	\begin{equation}
		w_k(z)=1+\frac{z^{1/2}}{g_k(z^{1/2})},\label{p5k}
	\end{equation}
	which is a PV transcendent for the parameters
	\begin{equation}
		a=\frac{(E_{0\ell}+\epsilon_1)^2}{2},\qquad b=-\frac{(E_{0\ell}-\epsilon_1+k-1)^2}{2},\qquad c=\frac{k-2E_{0\ell}}{2},\qquad d=-\frac{1}{8}.
	\end{equation}
	
\begin{figure}\centering
		\includegraphics[scale=0.37]{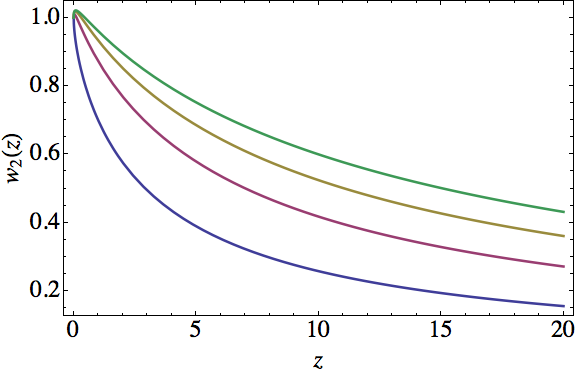}
		\includegraphics[scale=0.37]{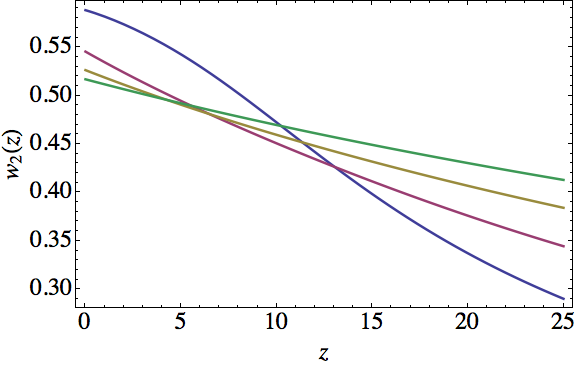}
		\caption{\small{PV solutions $w_2$ generated through second-order SUSY QM. The left plot is for the parameters $\ell=0$, $\nu_1=0$, and $\epsilon_1=\{1/4$ (blue), $-3/4$ (magenta), $-7/4$ (yellow), $-11/4$ (green)$\}$. The right plot is for $\epsilon_1=0$, $\nu_1=0$, and $\ell=\{$1 (blue), 3 (magenta), 6 (yellow), 10 (green)$\}$.}}\label{solskp5}
	\end{figure}

\begin{figure}\centering
		\includegraphics[scale=0.27]{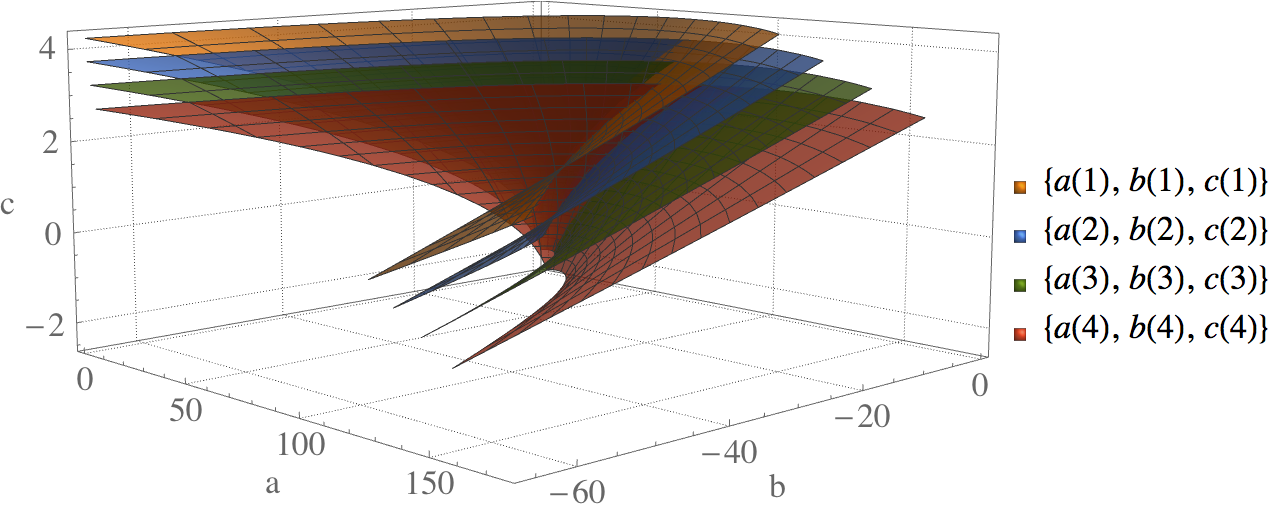}
		\caption{\small{Solution space for the PV equation generated through $k$th-order SUSY QM taking $\epsilon_1\in (-10,E_{0\ell})$ and $\ell\in (-1/2,10)$. We show the first four surface solutions for $k=\{1,2,3,4\}$ (yellow, blue, green and red, respectively).}}\label{paraRO2}
	\end{figure}

	In Figure~\ref{solskp5} we show some PV transcendents $w_2(z)$ obtained through the second-order SUSY transformation. Furthermore, with the $k$th-order SUSY QM we are able to expand the solution space $(a,b,c)$ by the inclusion of $k$. In Figure~\ref{paraRO2} we show a plot of such a solution space for $k=1,2,3,4$.
	
	\section{Complex solutions to the PV equation with real parameters}\label{cr}
	Let us stress that we can use the theorem of Section~\ref{sectmaRO} even with complex transformation functions. The simplest way to implement this is to use a complex linear combination of two standard linearly independent real solutions with a complex constant $\lambda+i\kappa$, with $\lambda, \kappa \in \mathbb{R}$, i.e.,
	\begin{align}
		u(x,\epsilon)=\, & x^{-\ell}\text{e}^{-x^2/4}\left[{}_1F_1\left(\frac{1-2\ell-4\epsilon}{4},\frac{1-2\ell}{2};\frac{x^2}{2}\right) \right.\nonumber\\
		&+\left.(\lambda+i\kappa)\left(\frac{x^2}{2}\right)^{\ell+1/2}{}_1F_1\left(\frac{3+2\ell-4\epsilon}{4},\frac{3+2\ell}{2};\frac{x^2}{2}\right)\right].\label{solRO2}
	\end{align}
	The result for the real case given in Equation~\eqref{solRO} is accomplished with the choice
	\begin{equation}
		\lambda= \frac{\Gamma\left(\frac{3+2\ell-4\epsilon}{4}\right)}{\Gamma\left(\frac{3+2\ell}{2}\right)}\nu,\qquad \kappa=0.
	\end{equation}
	
	Compared with the case when we were only dealing with real solutions that we studied in the previous section, the restriction $\epsilon_1\leq E_{0\ell}$ can now be surpassed, which implies that the solution space for the PV equation becomes even bigger for complex solutions than for real ones. In Table~\ref{tablepara} we show the parameters of the six solutions in terms of $\epsilon_1$, $\ell$ and $k$, which is the most general solution of the PV equation that we can find for this case.
	\begin{table}
		\begin{center}
			\begin{tabular}{cccc}
				\hline
				Index &$32a$&$32b$&$4c$\\
				\hline
				1234 & $(2\ell+4\epsilon_1+3)^2$ 	& $-(-2\ell+4\epsilon_1-4k+1)^2$ & $-2\ell+2k-3$\\
				1324 & $16 k^2$ 							& $-4(2\ell+1)^2$ 					& $4\epsilon_1-2k$\\
				1423 & $(-2\ell+4\epsilon_1+1)^2$ 	& $-(2\ell+4\epsilon_1-4k+3)^2$ 	& $2\ell+2k-1$\\
				2314 & $(2\ell+4\epsilon_1-4k+3)^2$ & $-(2\ell-4\epsilon_1-1)^2$ 		& $-2\ell-2k-1$\\
				2413 & $4 (2\ell+1)^2$					& $-16 k^2$ 							& $-4\epsilon_1+2k-4$\\
				3412 & $(2\ell-4\epsilon_1+4k-1)^2$& $-(2\ell+4\epsilon_1+3)^2$ 		& $2\ell-2k-1$\\
				\hline
			\end{tabular}
		\end{center}
		\vspace{-3mm}
		\caption{The six permutations of indexes of the extremal states which lead to different solutions to the PV equation due to the symmetries in the solution space.} \label{tablepara}
	\end{table}
	
\begin{figure}\centering
		\includegraphics[scale=0.37]{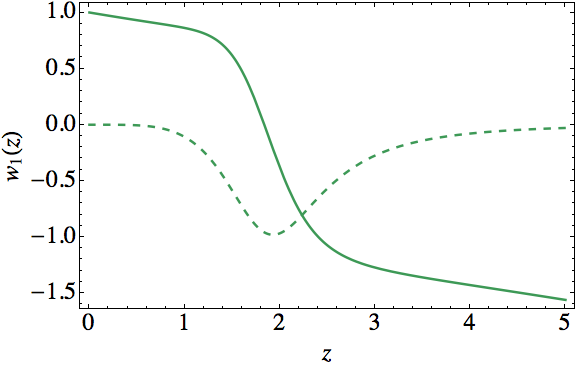}
		\includegraphics[scale=0.37]{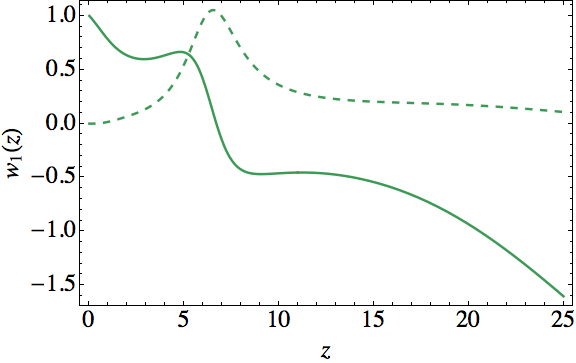}
		\caption{\small{Real (solid) and imaginary (dashed) parts of the solution $w_1(z)$ to PV equation for $\ell=3$, $\epsilon_1=0$, $\lambda=0$ and $\kappa=100$ (left) and $\ell=2$, $\epsilon_1=2$, $\lambda=0$ and $\kappa=\Gamma(-1/4)/\Gamma(7/4)$ (right).}}\label{comp5}
	\end{figure}

	The form of the solutions is the same as those of Equations~\eqref{VgkRO}, but now the linear combination in Equation \eqref{solRO2} is complex. In Figure~\ref{comp5} we show two complex PV transcendents associated with real parameters $a,b,c,d$.
	
	\section{Complex solutions to PV equation with complex parameters}\label{cc}
	We can also obtain complex solutions to the PV equation simply by allowing the factorization {\it energy} in Equation~\eqref{solRO2} to be complex. Then, as in the previous section, the solutions will also be complex but now the parameters $a,b,c$ of the PV equation will also be complex, as they depend on $\epsilon_1$.
	
\begin{figure}\centering
		\includegraphics[scale=0.37]{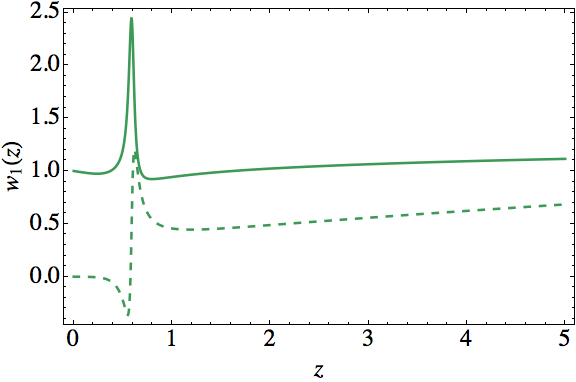}
		\includegraphics[scale=0.37]{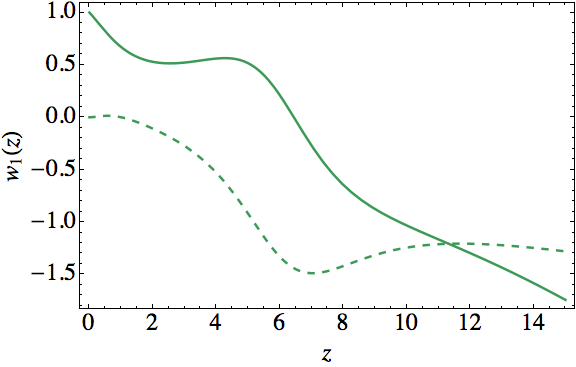}
		\caption{\small{Real (solid) and imaginary (dashed) parts of the solution $w_1(z)$ to PV equation for $\ell=3$, $\epsilon_1=1+11i$, and $\lambda +i\kappa=100i\ \Gamma(5/4-11i)/\Gamma(9/2)$; and $\ell=1$, $\epsilon_1=1-3i/5$, and $\lambda+i\kappa=(1-i)\Gamma(1/4+3i/5)/\Gamma(5/2)$.}}\label{comp52}
	\end{figure}

	For example, in Figure~\ref{comp52} we show two complex solutions to the PV equation associated with the complex parameters ($a,b,c$) given by
	\begin{subequations}
		\begin{alignat}{5}
			a&=-\frac{115}{4}+i\frac{429}{16}, & \qquad b& = \frac{1911}{32}+i\frac{55}{4}, & \qquad c &= \frac{49}{4},\\
			a&=-\frac{1881}{800}-i\frac{27}{20}, & \qquad b& = \frac{119}{800}-i\frac{3}{20}, & \qquad c &= -\frac{3}{4}.
		\end{alignat}
	\end{subequations}
	
	\section{Solution hierarchies for the PV equation}\label{hierarchies}
	The solutions $w(z)$ that we have found for the PV equation are expressed in terms of $g(x)$ in Equation~\eqref{p5k}, and at the same time $g(x)$ is expressed in terms of the functions $u_i$ in Equation~\eqref{VgkROb}. Recall that $u_i$ are formal eigenfunctions of the radial oscillator Hamiltonian with fixed eigenvalues, which are determined only by two complex parameters $\epsilon_1$ and $\lambda + i\kappa$. Also remember that all of them are eventually expressed in terms of the confluent hypergeometric function ${}_1F_1$, since $u_1$ is taken as in Equation~\eqref{solRO2}. Therefore, in the end our solutions to the PV equation will be written as functions depending on ${}_1F_1$.
	
	Recall that the Painlev\'e equations themselves define new {\it special functions}, the Painlev\'e transcendents, which are precisely the functions that solve the corresponding equation. Nevertheless, for some particular values of the parameters $\epsilon_1$ and $\lambda+i\kappa$, they can be expressed in terms of known special functions. This is useful to define {\it solution hierarchies}, which has been done previously for the PIV equation \cite{Ber13,BF13a,BCH95}. Next, we will do a similar classification for the PV transcendents $w(z)$. Most of them will correspond to the broad category of \textit{rational solutions}, expressed as fractions of polynomials. Let us note that most of the members of the rational hierarchy are generated by the same SUSY transformations that give place to the \emph{exceptional orthogonal polynomials}, which have recently received a lot of attention \cite{GKM10,STZ10,OS11,MQ13,SR13,GKM14,Que15}.
	
	\subsection{Laguerre polynomials hierarchy}
	When one of the following two conditions is fulfilled
	\begin{subequations}
		\begin{alignat}{3}
			\epsilon_1= & \ n-\frac{\ell}{2}+\frac{1}{4}, & \qquad \nu_1 & =0, \\
			\epsilon_1= & \ n+\frac{\ell}{2}+\frac{3}{4}, & \qquad \nu_1 & \rightarrow\infty,
		\end{alignat}
	\end{subequations}
	the confluent hypergeometric function reduces to a Laguerre polynomial due to the following identity
	\begin{equation}
		L_n^{\alpha}(x)=\frac{(\alpha+1)_n}{n!}{}_1F_1(-n,\alpha+1,x).
	\end{equation}
	Two examples of solutions to the PV equation belonging to this hierarchy are
	\begin{subequations}
		\begin{align}
			w_1(z)=& \ 1-z^{-1/2},\\
			w_1(z)=& \  1-\frac{z^{3/2}L_1^{(\alpha)}(z^2/2)}{2L_1^{(\alpha)}(z^2/2)-2\alpha-1},
		\end{align}
	\end{subequations}
	where $\alpha=-(2\ell+1)/2$.
	
	\subsection{Hermite polynomials hierarchy}
\begin{figure}\centering
		\includegraphics[scale=0.37]{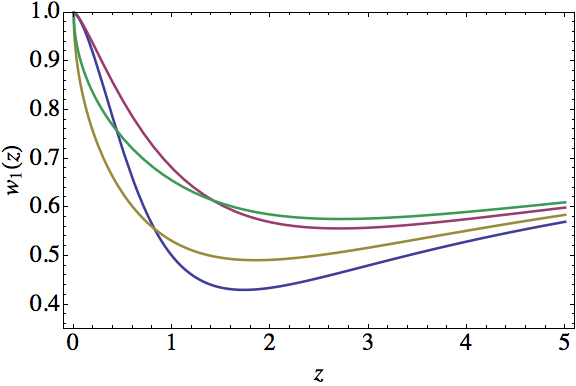}
		\caption{\small{Solutions $w_1(z)$ belonging to the Hermite polynomials hierarchy given by Equation \eqref{w1hermitea} for $n=0,1$ (blue, magenta) and by \eqref{w1hermiteb} for $n=1,2$ (yellow, green).}}\label{w1her}
	\end{figure}

	Take now one of
	\begin{subequations}
		\begin{alignat}{5}
			\ell & =0, & \qquad \epsilon_1 & =n+1/4, & \qquad \nu_1 & =0,\\
			\ell & =0, & \qquad \epsilon_1 & =n+3/4, & \qquad \nu_1 & \rightarrow\infty.
		\end{alignat}
	\end{subequations}
	We obtain then the Hermite polynomial hierarchy. Two examples of PV transcendents belonging to this hierarchy are
	\begin{subequations}\begin{align}
			w_1(z)&=1-\frac{z^{3/2}H_{2n}(z)}{(z^2+1)H_{2n}(z)-4nzH_{2n-1}(z)},\label{w1hermitea}\\
			w_1(z)&=1+\frac{z^{1/2}H_{2n}(z)}{4nH_{2n-1}(z)-zH_{2n}(z)}\label{w1hermiteb}.
		\end{align}\label{w1hermite}\end{subequations}
	\hspace{-1mm}where $H_n(x)$ are the Hermite polynomials. In Figure~\ref{w1her} we have plotted two members of each of these two solution families for two values of $n$. In the plots, it looks like the solutions may present a singularity at $x=0$, but this indeed does not happen, as can be proven analytically using Equations~\eqref{w1hermite}.
	
	\subsection{Weber or Parabolic cylinder hierarchy}
	In order to reduce the confluent hypergeometric function ${}_1F_1$ into a Weber or parabolic cylinder function $E_\mu (x)$, the following conditions must be fulfilled:
	\begin{equation}
		\ell=0,  \qquad \epsilon_1  =(2\mu+1)/4, \qquad \nu_1 =0.
	\end{equation}
	Two examples of solutions $w_1(z)$ of this hierarchy that are obtained through 1-SUSY are
	\begin{subequations}
		\begin{align}
			w_1(z)&=1-\frac{2z^{3/2}E_\mu(z)}{2(z^2+1)E_\mu(z)-zE_{\mu-1}(z)+zE_{\mu+1}(z)},\\
			w_1(z)&=1-\frac{2z^{1/2}E_\mu(z)}{2(z^2+1)E_\mu(z)-zE_{\mu-1}(z)+zE_{\mu+1}(z)}.
		\end{align}
	\end{subequations}
	
	\subsection{Modified Bessel hierarchy}
\begin{figure}
		\begin{center}
			\includegraphics[scale=0.37]{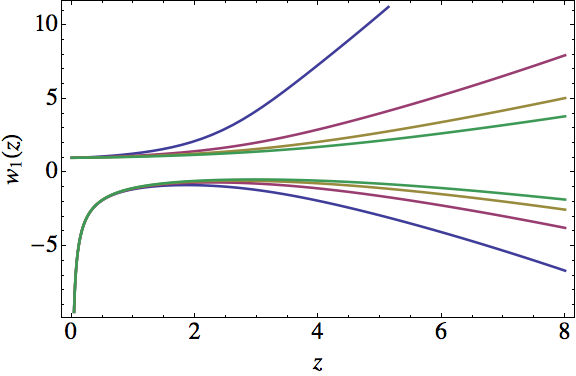}
			\caption{\small{Solutions $w_1(z)$ of the PV equation given by Equations~\eqref{w1bessel} which belong to the modified Bessel hierarchy. The positive solutions belong the the first family and negative solutions to the second one for $\mu=1$ (blue), $\mu=2$ (magenta), $\mu=3$ (yellow), and $\mu=4$ (green).}}\label{w1bes}
		\end{center}
	\end{figure}

	Under the conditions
	\begin{subequations}
		\begin{alignat}{5}
			\ell & =-(4\mu+1)/2, & \qquad \epsilon_1 & =0, & \qquad \nu_1 & =0,\\
			\ell & =-(4\mu +3)/2, & \qquad \epsilon_1 & =0, & \qquad \nu_1 & =0,\\
			\ell & =(4\mu-1)/2, & \qquad \epsilon_1 & =0, & \qquad \nu_1 & \rightarrow\infty,\\
			\ell & =(4\mu+1)/2, & \qquad \epsilon_1 & =0, & \qquad \nu_1 & \rightarrow\infty,
		\end{alignat}
	\end{subequations}
	the function ${}_1F_1$ reduces to the modified Bessel function $I_\mu(z)$. Two examples of the corresponding PV transcendents are
	\begin{subequations}
		\begin{align}
			w_1(z)&=1-\frac{2z^{3/2}I_\mu (z^2/4)}{(z^2-8\mu)I_\mu (z^2/4)-z^2I_{\mu+1}(z^2/4)},\\
			w_1(z)&=1+\frac{2I_\mu (z^2/4)}{z^{1/2}[I_{\mu+1}(z^2/4)-I_{\mu}(z^2/4)]}.
		\end{align}\label{w1bessel}
	\end{subequations}
	\hspace{-1.5mm}We present four examples for each of the two solution families of Equations \eqref{w1bessel} in Figure \ref{w1bes}.
	
	\subsection{Exponential hierarchy}
\begin{figure}\centering
		\includegraphics[scale=0.37]{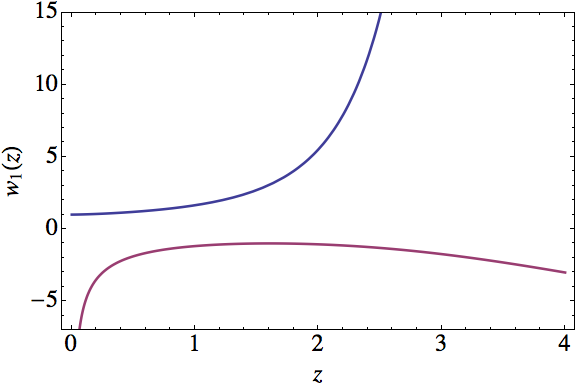}
		\caption{\small{The two PV transcendents $w_1(z)$ given by Equations~\eqref{w1expsol} belonging to the exponential hierarchy.}}\label{w1exp}
	\end{figure}

	For special values of the parameters of the SUSY transformation, the confluent hypergeometric functions reduces to a polynomial, but there is still the exponential function as a factor in the general solution $u_1(x)$, which could appear in the PV transcendent. The conditions
	\begin{equation}
		\ell =1/2, \qquad \epsilon_1 =0,  \qquad \nu_1 \rightarrow\infty,
	\end{equation}
	illustrate this situation, with two solutions which are obtained through $1$-SUSY given by
	\begin{subequations}
		\begin{align}
			w_1(z)&=1+\frac{\exp(z^2/2)-1}{z^{1/2}},\\
			w_1(z)&=1-\frac{z^{3/2}}{2}+\frac{z^{7/2}}{2z^{2}+4-4\exp(z^2/2)}.
		\end{align}\label{w1expsol}
	\end{subequations}
	\hspace{-1mm}In Figure \ref{w1exp} we show these two solutions, which belong to the exponential hierarchy.
	
	\subsection{Polynomial hierarchy not from special functions}
	Usually, there are some conditions for the ${}_1F_1$ to reduce to an exponential. In such a case, this exponential together with the one of the general solution $u_1(x)$ at the end causes the PV solution to be a polynomial, but different from any other hierarchy. Two examples are obtained for
	\begin{subequations}
		\begin{alignat}{3}
			\epsilon_1 & =\frac{\ell}{2}-\frac{1}{4}, & \qquad \nu_1 & =0,\\ 
			\epsilon_1 & =-\frac{\ell}{2}-\frac{3}{4}, & \qquad \nu_1 & \rightarrow\infty.
		\end{alignat}
	\end{subequations}
	An explicit PV transcendent of this hierarchy is given by
	\begin{equation}
		w_1(z)=1-\frac{z^{3/2}}{2\ell+1}.
	\end{equation}
	
	\section{Conclusions}\label{conclusions}
	In this paper we presented an algebraic technique to solve the PV equation. This method supplies specific subsets of PV transcendents that are connected with the confluent hypergeometric function and other related functions. In order to introduce our procedure, we revisited the SUSY QM technique with special emphasis in the radial oscillator in Section \ref{capsusyqm} and the polynomial Heisenberg algebras in Section \ref{pha}, particularly the third-order PHA and the related fourth-order ladder operators in Section \ref{topha}. In the last case, we established the connection of this algebra with solutions of the PV equation, through four first-order intertwining operators.
	
	Later, we derived the SUSY partners of the radial oscillator and we obtained the natural ladder operators for these systems in Section \ref{SUSYRO}. Then, in Section \ref{sectmaRO} we proved a reduction theorem, in which we pointed out the requirements for the $k$th-order SUSY partners of the radial oscillator, which usually have $(2k+2)$th-order ladder operators, to be described as well by reduced fourth-order ones and therefore related with third-order PHA and the PV equation.
	
	Using the connection between SUSY partners of the radial oscillator and third-order PHA, in Section \ref{solsPV} we were able to implement a method to obtain solutions of the PV equation, i.e., PV transcendents. We presented the general method for the higher-order SUSY partners of the radial oscillator and we also worked out explicit solutions coming from the radial oscillator and its first- and second-order SUSY partners. Then, we obtained real and complex solutions associated with real parameters $a,b,c,d$ of the PV equation in Sections \ref{rr} and \ref{cr}, respectively, as well as complex solutions for complex parameters in Section \ref{cc}.
	
	Finally, in Section \ref{hierarchies} we classified some of the PV transcendents obtained into solution hierarchies, according to the special functions they depend on, e.g., Laguerre polynomials, Hermite polynomials, Weber functions, exponentials, among others.
	
	This work further expands the previous treatment of the authors for the Painlev\'e IV equation \cite{Ber10,BF11a,Ber13,BF11b,Ber12,BF12,BF13a}, with the main difference that the connection with the PV equation is more elaborated, and working out the explicit solutions becomes increasingly complicated even at lower orders. We realize also that the classification done here is probably still incomplete, as there are even further solutions that can be obtained by this method. 
	
	The relation between the PV equation and SUSY QM has other potential applications. Let us mention that some of the fourth-order ladder operators obtained by the reduction theorem describe a kind of exceptional polynomials. Another example is that, in the context of SUSY QM, Darboux transformations can be designed to produce a kind of B\"acklund transformations for different Painlev\'e V equations. We will continue working in this direction in the near future.

\section*{Acknowledgement}
DB and DJFC acknowledge the financial support of Conacyt (Mexico) project 152574.
JN acknowledges partial financial support to the Spanish MINECO (Project MTM2014-57129-C2-1-P) and Junta de Castilla y Le\'on (VA057U16).
DB also acknowledges the support of Universidad de Valladolid.

\bibliographystyle{ieeetr}
\bibliography{references}
\end{document}